\documentclass[lettersize,journal]{IEEEtran}
\usepackage{amsmath,amsfonts}
\usepackage{amssymb,amsthm,amscd}
\usepackage{mathtools}
\usepackage{commath,mathabx}
\usepackage{algorithmic}
\usepackage{algorithm}
\usepackage{array}
\usepackage{textcomp}
\usepackage{multirow}
\usepackage{stfloats}
\usepackage{cancel}
\usepackage[labelformat=simple]{subcaption}
\usepackage[overload]{empheq}
\usepackage{psfrag,color}
\usepackage{url}
\usepackage{verbatim}
\usepackage{tabstackengine}
\usepackage{graphicx}
\usepackage[sort]{cite}
\usepackage{tabularx}
\usepackage{blkarray}
\newcommand{\T}{{\scalebox{.65}{$\rm T$}}}

\newcommand{\BM}{{\mathbf B}}
\newcommand{\CM}{{\mathbf C}}

\newcommand{\GAM}{{\boldsymbol \Gamma}}

\newcommand{\IM}{{\mathbf I}}

\newcommand{\RM}{{\mathbf R}}

\newcommand{\E}{{\rm E}}
\newcommand{\uM}{{\mathbf u}}

\newcommand{\bM}{{\mathbf b}}

\newcommand{\w}{{\mathbf w}}
\newcommand{\psiM}{{\boldsymbol \psi}}

\newcommand{\wo}{{\mathbf w}_{\rm o}}
\newcommand{\wtil}{\widetilde{\mathbf{w}}}

\newcommand{\Tr}{{\rm Tr}}

\definecolor{laranja}{rgb}{0.8,0.5,0}
\definecolor{capri}{rgb}{0.0, 0.75, 1.0}
\definecolor{carmine}{rgb}{0.59, 0.0, 0.09}
\definecolor{dimgray}{rgb}{0.41, 0.41, 0.41}

\newcommand{\vect}{\mathrm{vec}}

\DeclareMathAlphabet\mathbfcal{OMS}{cmsy}{b}{n}
\DeclareMathAlphabet\mathbfcal{OMS}{cmsy}{b}{n}
\def\code#1{\texttt{#1}}

\newcommand{\boxedeqn}[1]{
	\begin{equation}
		\fbox{$\displaystyle #1 $}
\end{equation}}
\newtheorem{theorem}{Result}

\begin{document}

\title{{On the Impact of Random Node Sampling on Adaptive Diffusion Networks}}

\author{Daniel G. Tiglea, Renato Candido, and Magno T.~M.~Silva,~\IEEEmembership{Member, IEEE} \thanks{This work was supported by CAPES under Grant 88887.512247/2020-00 and Finance Code 001, by CNPq under Grant 303826/2022-3, and by FAPESP under Grant 2021/02063-6. The authors are with the Electronic Systems Engineering Department, Escola Politécnica, University of São Paulo, São Paulo, SP, Brazil, e-mails:\{dtiglea,~renatocan,~magno\}@lps.usp.br, ph. +55-11-3091-5134.}}

\markboth{Submitted to the IEEE Transactions on Signal Processing}%
{Shell \MakeLowercase{\textit{et al.}}: A Sample Article Using IEEEtran.cls for IEEE Journals}


\maketitle

\begin{abstract}
In this paper, we analyze the effects of random sampling on adaptive diffusion networks. These networks consist in a collection of nodes that can measure and process data, and that can communicate with each other to pursue a common goal of estimating an unknown system. In particular, we consider in our theoretical analysis the diffusion least-mean-squares algorithm in a scenario in which the nodes are randomly sampled. Hence, each node may or may not adapt its local estimate at a certain iteration. Our model shows that, if the nodes cooperate, {a reduction in the sampling probability} leads to a slight {decrease in the steady-state Network Mean-Square Deviation (NMSD), assuming that the environment is stationary and that all other parameters of the algorithm are kept fixed.} Furthermore, under certain circumstances, this can also ensure the stability of the algorithm in situations in which it would otherwise be unstable. Although counter-intuitive, our findings are backed by simulation results, which match the theoretical curves well.
\end{abstract}


\section{Introduction} \label{sec:intro}
\IEEEPARstart{A}{daptive} diffusion networks have attracted widespread attention in the distributed signal processing literature over the past decade and a half, and have become consolidated tools in the area~\cite{Sayed_Networks2014,Sayed_capitulo2014,Sayed_Proc2014,lopes2008diffusion, cattivelli2009diffusion,cattivelli2008diffusion,li2009distributed,Fernandez-Bes2017,NassifICASSP2018,HuaEusipco2018}. They consist in a set of connected \emph{agents}, or \emph{nodes}, that are able to measure and process data locally, and that can communicate with other nodes in their vicinity. The network formed by these agents has a collective goal to estimate a parameter vector of interest, without the need for a central processing unit~\cite{Sayed_Networks2014,Sayed_capitulo2014,Sayed_Proc2014,lopes2008diffusion, cattivelli2009diffusion,cattivelli2008diffusion,li2009distributed,Fernandez-Bes2017,NassifICASSP2018,HuaEusipco2018}. Hence, they present better flexibility, scalability, and robustness than centralized approaches, whose central unit represents a critical point of failure, and limits the area where the nodes can be deployed.

In order to enable the distributed learning of the parameters, each node usually computes its own \emph{local} estimate in what is known as the \emph{adaptation step}. Then, the neighboring nodes cooperate to reach a \emph{global} estimate of the vector of interest. This stage is usually referred to as the \emph{combination step}~\cite{Sayed_Networks2014,Sayed_capitulo2014,Sayed_Proc2014,lopes2008diffusion, cattivelli2009diffusion,cattivelli2008diffusion,li2009distributed,di2013bio,Fernandez-Bes2017,NassifICASSP2018,HuaEusipco2018}. Due to their advantages in comparison with other distributed settings, such as the incremental~\cite{Cassio06,Cassio2007} and consensus~\cite{olfati2005consensus,spanos2005dynamic} strategies, adaptive diffusion networks have branched out into many different research topics. Examples include multitask networks~\cite{chen2014multitask,chen2015diffusion,plata2015distributed,gogineni2021performance}, nonlinear adaptive networks~\cite{gao2015diffusion,chouvardas2016diffusion,shin2018distributed,bouboulis2018random}, among others. Moreover, the field of graph signal processing (GSP) has drawn inspiration from these techniques, since its applications are usually distributed in nature~\cite{NassifICASSP2018,HuaEusipco2018,DiLorenzoTSP2017,di2017distributed}. Hence, many graph adaptive filtering algorithms can be seen as an extension of adaptive diffusion networks to domains where space, as well as time, plays a role in the development of the signals of interest~\cite{tiglea2020low}.

For the feasibility of these solutions, it is oftentimes desirable to restrict the amount of data measured and processed by the nodes. This process has been named as \textit{sampling} in the literature~\cite{DiLorenzoTSP2018,DiLorenzoTSP2017}. By sampling only some of the nodes at each iteration, we can reduce the computational and memory burdens associated with the learning task. However, there may also be a negative impact on the convergence rate. Based on these observations, in~\cite{tiglea2020low}, we proposed an adaptive sampling algorithm for adaptive diffusion networks. Later, some modifications to this algorithm were proposed in~\cite{tiglea2022adaptive}. In both papers, we adopted a random sampling technique as a benchmark to compare the proposed solutions with, and observed a peculiar phenomenon. When using the random sampling method, the convergence of the network deteriorates as we decrease the number of nodes sampled per iteration, as one might expect. What is interesting, however, is that {apparently} the steady-state performance slightly improves as we sample less nodes in stationary environments. Next, we provide some preliminary simulation results to illustrate this. The discussion on the results presented will help us motivate the present work. 

\subsection{Introductory Simulations and Motivation} \label{sec:intro_simu}

\begin{figure}[htb!]
	\centering
	\includegraphics*[trim=0.15cm 0.2cm 0.15cm 0.1cm, clip=true,width=1\columnwidth]{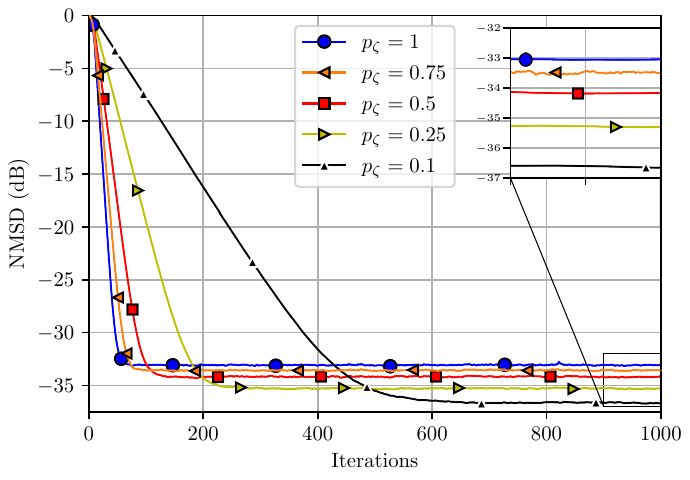}
	\caption{\protect\label{fig:intro} Performance of the ATC dLMS algorithm with a random sampling technique, considering different sampling probabilities $p_{\zeta}$, in the Scenario 1 described in Sec.~\ref{sec:simulation}.}
	
\end{figure}

We show in Fig.~\ref{fig:intro} simulation results obtained with the adapt-then-combine diffuse Least-Mean-Squares (ATC dLMS) algorithm~\cite{Sayed_Networks2014,Sayed_capitulo2014,Sayed_Proc2014,lopes2008diffusion, cattivelli2009diffusion} with a random sampling technique. In this setup, each node is sampled with probability $p_{\zeta}$, or not sampled with probability $1 - p_{\zeta}$. This algorithm will be revisited in detail in Sec.~\ref{sec:problem}. We consider the Scenario 1 described in Sec.~\ref{sec:simulation}, and compare the behavior of the algorithm for different values of $p_{\zeta} \in \{0.1,0.25,0.5,0.75,1\}$. As a performance indicator, we adopt the Network Mean-Square Deviation (NMSD), defined in Sec.~\ref{sec:analysis}.

From Fig.~\ref{fig:intro}, we can clearly see that the smaller the sampling probability, the slower the convergence rate. {Interestingly, however}, the steady-state NMSD decreases as we reduce {the sampling probability. In relation to the case in which every node is sampled, i.e., $p_\zeta = 1$, the difference in steady-state NMSD reaches approximately 3.5 dB, when we examine the curve obtained with $p_\zeta = 0.1$.} This observation is especially intriguing, as it seems counter-intuitive. After all, we {observe a reduction in the steady-state NMSD while using less data, rather than more}, in a completely random fashion. {Thus, our goal in this paper is to investigate the effects of random sampling on adaptive diffusion networks}. 

\subsection{Relations with Other Works and Major Contributions}

To the best of our knowledge, this phenomenon has not been pointed out in the literature. For example, in~\cite{tu2012influence,sayed2013online,sayed2013diffusion}, a scenario was considered in which some of the nodes are not capable of performing the adaptation step. These are referred to as ``uninformed'' nodes, in contrast with the ``informed'' nodes, which can perform it. In those works, it was shown that, in comparison with a network in which every node is informed, the steady-state NMSD can decrease, increase, or remain unchanged, as we turn some of the nodes into uninformed ones. However, differently from the scenario considered in the simulations of Fig.~\ref{fig:intro}, informed nodes carry out the adaptation step at every time instant, whereas uninformed nodes never do so. Moreover, in this case, the enhancement or the deterioration in the steady-state NMSD depends on the noise power at the informed or uninformed nodes. In other words, prior knowledge of the noise variance at each node is required. This differs with the behavior observed in Fig.~\ref{fig:intro}, in which the nodes are sampled randomly, and this still leads to a decrease in the steady-state NMSD. Furthermore, in~\cite{zhao2014asynchronous,zhao2014asynchronous2,zhao2014asynchronous3}, the authors study networks in which the adaptation and combination steps are not necessarily carried out simultaneously by all the nodes at every iteration. These networks are referred to as ``asynchronous'', in contrast with the ``synchronous'' ones that appear, e.g., in~\cite{Sayed_Networks2014,Sayed_capitulo2014,Sayed_Proc2014,lopes2008diffusion, cattivelli2009diffusion}. From this perspective, the random-sampling network used in the simulations of Fig.~\ref{fig:intro} can be deemed as a type of asynchronous network. However, in those papers, the phenomena seen in Fig.~\ref{fig:intro} were not observed.

{Next, we provide a list of the most relevant contributions of this paper in comparison with previous works:}
\begin{itemize}
\item {We obtain theoretical results that describe the effects of the random sampling of the nodes on the transient and steady-state behaviors of the dLMS algorithm observed in Fig.~\ref{fig:intro}. Furthermore, and perhaps more importantly, our analysis helps us explain \textit{why} these effects occur, and to what we may attribute them;}
\item  {Different from the studies conducted in~\cite{zhao2014asynchronous,zhao2014asynchronous2,zhao2014asynchronous3}, which are based on the Energy Conservation Argument -- a powerful tool for the analysis of adaptive algorithms~\cite{SayedL2008} --, ours analysis utilizes a traditional statistical framework~\cite{Haykin_AFT2014,CapituloVitor}, which facilitates the interpretation of some results;}
\item {Several simulation results are presented considering different scenarios, therefore abundantly illustrating the phenomena of interest and the theoretical results. In particular, we present simulation results that showcase the effects of the sampling of the nodes on the stability of the algorithm, which depends on the selection of the combination weights and on the network topology;}
\item {We study the impact of the random node sampling on the computational cost.}
\end{itemize}

In addition to the insights that this study brings into the inner workings of these solutions, we believe that this is a matter of practical interest, as it can aid in the development of efficient algorithms for adaptive diffusion networks. {We remark that our goal in this paper is not to motivate the usage of the random sampling technique of Fig.~\ref{fig:intro}, but just to study the effects of the sampling of the nodes. For instance, if we manage to keep the nodes sampled in the transient phase, and cease to sample them in steady state, we} could mitigate the deterioration in the convergence rate observed in Fig.~\ref{fig:intro}, while {potentially} reducing the computational cost and improving the performance in steady state {in comparison with the case in which all the nodes are sampled at every iteration}. That is the goal of, e.g., the algorithms of~\cite{tiglea2020low} and~\cite{tiglea2022adaptive}, but other solutions could be proposed in the future. This approach simultaneously promotes the feasibility of adaptive diffusion networks in practice, and enhances their performance, which shows how promising this topic is for future research. 

\subsection{Organization of the Paper and Notation} \label{sec:organization}

The remainder of this paper is organized as follows. In Sec.~\ref{sec:problem}, we present the problem formulation. In Sec.~\ref{sec:analysis}, we conduct a theoretical analysis on the behavior of adaptive diffusion networks, and on the effects of sampling on them. In Sec.~\ref{sec:computational}, the impact of sampling on the computational cost is analyzed. Finally, in Secs.~\ref{sec:simulation} and~\ref{sec:conclusions}, we present simulation results and the main conclusions of our work, respectively.

\noindent \textbf{Notation}. We use normal font letters for scalars, boldface lowercase letters for vectors, and boldface uppercase letters for matrices. Moreover, $(\cdot)^{\T}$ denotes transposition, $\E\{\cdot \}$ the mathematical expectation, $[\cdot]_{\ell k}$ the element of a matrix located at its $\ell$-th row and $k$-th column, $\Tr[\cdot]$ the trace of a matrix, $\delta_{ij}$ the Kronecker delta function of $i$ and $j$, $\|\cdot\|$ the Euclidean norm, $|\cdot|$ the absolute value, $\slash$ the difference between two sets, $\otimes$ the Kronecker product, and $\odot$ the Hadamard product. We denote the $L \times L$ identity matrix by $\IM_L$, an $L$ by $M$ matrix of zeros by $\mathbf{0}_{L \times M}$, and an $L$ by $M$ matrix of ones by $\mathbf{1}_{L \times M}$. When referring to $L$-length column vectors of zeros or ones, we write $\mathbf{0}_L$ and $\mathbf{1}_L$, respectively.  Finally, we denote by $\vect\{\cdot\}$ the vectorization of a matrix by stacking all of its columns together to form a single column vector. To simplify the arguments, we assume real data throughout the paper.

\section{Problem Formulation} \label{sec:problem}

Let us consider a network consisting of $V$ nodes, with labels $k \!\in\! \{1, \cdots, V\}$. For each node $k$, we call the set of nodes with which it can communicate, including node $k$ itself, its neighborhood, and we denote it by $\mathcal{N}_k$. We assume that each node $k$ has access at each iteration $n$ to an input signal $u_k(n)$ and to a desired signal $d_k(n)$, which we model as\mbox{\cite{Sayed_Networks2014,Sayed_capitulo2014,Sayed_Proc2014,lopes2008diffusion, cattivelli2009diffusion}}
\begin{equation} \label{eq:dk}
	d_k(n) = \uM_k^{\T}(n)\mathbf{w}^{\rm o} + v_k(n), 
\end{equation}
{where $\mathbf{w}^{\mathrm{o}}$ is an $M$-length column vector that represents an unknown system and $\uM_k(n) = [u_k(n)\ u_k(n-1) \ \cdots\  u_k(n-M+1)]^{\rm T}$ is a regressor vector formed by the last $M$ samples of the input signal. Finally, $v_k(n)$ is the measurement noise at node $k$. The vector $\mathbf{w}^{\mathrm{o}}$ is oftentimes referred to as the optimal system in the adaptive filtering literature~\cite{Sayed_Networks2014,Sayed_capitulo2014,Sayed_Proc2014,lopes2008diffusion, cattivelli2009diffusion}}. Throughout this paper, we assume that the optimal system is the same for all the nodes in the network. Adaptive diffusion networks proposed for this type of scenario are oftentimes called ``single-task'', in opposition to the multitask networks that emerge when the optimal system can be different for each node~\cite{chen2014multitask,chen2015diffusion,plata2015distributed,gogineni2021performance}. It is worth noting that~\eqref{eq:dk} can also be used to model processes in GSP~\cite{NassifICASSP2018,HuaEusipco2018}, although in this case $\uM_k(n)$ is no longer construed as a regressor vector, but instead gathers information from the input signal at the neighbors of node $k$~\cite{tiglea2020low}.


The goal of the network is to obtain an estimate $\w$ of $\mathbf{w}^\mathrm{o}$ in a distributed manner by solving~\cite{Sayed_Networks2014,Sayed_capitulo2014,Sayed_Proc2014,lopes2008diffusion, cattivelli2009diffusion}
\begin{equation} \label{eq:minimize}
	\min_{\mathbf{w}} J_{\text{global}}(\w)\!=\!\min_{\mathbf{w}}\sum_{k=1}^{V}J_k(\w),
\end{equation}
where $J_k(\w)$ are local cost functions at each node $k$. A common choice for the $J_k(\w)$, $k=1,\cdots,V$ is the mean squared error (MSE)~\cite{SayedL2008,Haykin_AFT2014}, {in which case each node $k$ should seek, at every iteration $n$, to minimize}~\cite{Sayed_Networks2014,Sayed_capitulo2014,Sayed_Proc2014,lopes2008diffusion, cattivelli2009diffusion}
\begin{equation} \label{eq:cost_w}
	J_k(\w)\!\triangleq\!\E\{[d_k(n)\!-\!\uM_k^{\T}(n)\w]^2\}.
\end{equation}
{To this end, each} node $k$ computes a local estimate of $\w^{\rm o}$ in order to minimize its individual cost function $J_k(\w)$. {For this purpose}, it uses the data available locally, as well as the estimates by neighboring nodes. This is known as the adaptation step. Then, the node $k$ cooperates with its neighbors to form a combined estimate in what is called the combination step. Depending on how the adaptation of the local estimates is carried out, different algorithms can be obtained. Examples include the dLMS algorithm~\cite{Sayed_Networks2014,Sayed_capitulo2014,Sayed_Proc2014,lopes2008diffusion, cattivelli2009diffusion}, the diffusion recursive least squares (dRLS)~\cite{cattivelli2008diffusion}, diffusion Normalized LMS (dNLMS)~\cite{lopes2008topologies,tiglea2020low,tiglea2022adaptive}, diffusion Affine Projection Algorithm (dAPA)~\cite{li2009distributed}, among others~\cite{lu2017diffusion,werner2009distributed}. For the sake of simplicity, we shall focus our analysis in this paper on the dLMS algorithm, which is obtained by adopting an LMS-type of strategy~\cite{SayedL2008,Haykin_AFT2014} for the update of the local estimates. Considering a scenario in which each node $k$ may or may not update its local estimate at each iteration $n$, the adaptation and combination steps of the dLMS algorithm are respectively given by~\cite{Sayed_Networks2014,Sayed_capitulo2014,Sayed_Proc2014,lopes2008diffusion, cattivelli2009diffusion}
\begin{subequations} \label{eq:atc_dlms}
	\begin{empheq}[left={\empheqlbrace\,}]{align}
		&\psiM_k(n)\!=\!\w_k(n-1)\!+\!\mu_k \zeta_k(n) \uM_k(n) e_k(n) \label{eq:atc_dlms1} \\		
		&\w_k(n)=\sum_{i \in \mathcal{N}_k} c_{i k} \psiM_i(n) \label{eq:atc_dlms2}.
	\end{empheq}
\end{subequations}
where {$\psiM_k$ and $\w_k$ are the local and combined estimates of $\wo$ at node $k$, respectively, $\mu_k > 0$ is a step size, 
\begin{equation}\label{eq:ek}
	e_k(n) = d_k(n) - \uM_k^{\rm T}(n) \w_k(n-1)
\end{equation}
is the estimation error at node $k$, and $\zeta_k(n) \!\in\! \{0,1\}$ is a binary variable such that $\zeta_k(n) \!=\! 1$ if node $k$ is sampled, and $\zeta_k(n) \!=\! 0$ otherwise. In the former case, $\psiM_k(n)$ is updated as usual in the dLMS algorithm~\cite{Sayed_Networks2014,Sayed_capitulo2014,Sayed_Proc2014,lopes2008diffusion, cattivelli2009diffusion}. In contrast, when node $k$ is not sampled, $\uM^{\rm T}_k(n) \w_k(n-1)$ and $e_k(n)$ do not need to be computed, as~\eqref{eq:atc_dlms1} becomes simply $\psiM_k(n) \!=\! \w_k(n-1)$. Lastly, $c_{ik}$ are combination weights satisfying\mbox{\cite{Sayed_Networks2014,Sayed_capitulo2014,Sayed_Proc2014,lopes2008diffusion, cattivelli2009diffusion}}
\begin{equation} \label{eq:ck}
	\!c_{ik}(n) \!\geq\!0,\ \!\textstyle\sum_{i\in \mathcal{N}_k}\!c_{ik}(n)\!=\!1,\ \text{and}\ c_{ik}(n)\!=\!0 \ \text{for}\ i\!\notin\! \mathcal{N}_k.
\end{equation}
There are several possible rules for the selection of the combination weights. For instance, if we adopt $c_{ik} = 1$ if $i = k$ and $c_{ik} = 0$ otherwise, this corresponds to a setup in which the nodes do not exchange their local estimates. This is oftentimes referred to as the non-cooperative approach in the literature~\cite{Sayed_Networks2014,Sayed_capitulo2014,Sayed_Proc2014,lopes2008diffusion, cattivelli2009diffusion}, and can be seen as a scenario in which $V$ adaptive filters try to solve~\eqref{eq:minimize} isolated from each other. Cooperative strategies include the Uniform, Metropolis, and Hastings rules, among others~\cite{Sayed_Networks2014,Sayed_capitulo2014,Sayed_Proc2014}. Moreover, several adaptive
schemes have been proposed in the literature~\cite{Fernandez-Bes2017,takahashi2010diffusion,yu2013strategy,tu2011optimal}, in which the combination weights $c_{ik}(n)$ are adjusted along the iterations. For simplicity, in our analysis we only consider static combination weights. In Table~\ref{tab:ck}, we provide a summary of the static rules considered in the simulations.

\begin{table}[h!tb]
	\centering
	\caption{Summary of some static rules for the selection of the combination weights most widely adopted in the literature.} \label{tab:ck}
	\begin{tabularx}{1\columnwidth}{|p{2.4cm}|X|}
		\hline
		Name & Equations \\
		\hline
		\!\!\! Non-coop.~\cite{Sayed_Networks2014,Sayed_capitulo2014,Sayed_Proc2014,lopes2008diffusion, cattivelli2009diffusion} & $\!\!\!c_{i k}\!=\! 
		\begin{cases}
			1,\ \text{if}\ i=k \\
			0, \ \hspace*{0.55cm} \text{otherwise}
		\end{cases}$ \\
		\hline
		\!\!\! Uniform~\cite{Sayed_Networks2014,Sayed_capitulo2014,Sayed_Proc2014,lopes2008diffusion, cattivelli2009diffusion} & $\!\!\!c_{i k}\!=\! 
		\begin{cases}
			\dfrac{1}{|\mathcal{N}_k|},\ \text{if}\ i \in \mathcal{N}_k \\
			0, \ \hspace*{0.55cm} \text{otherwise}
		\end{cases}$ \\
		\hline
		\!\!\! Metropolis~\cite{Sayed_Networks2014,Sayed_capitulo2014,Sayed_Proc2014,lopes2008diffusion, cattivelli2009diffusion}  &  $\!\!\!	c_{i k}\!=\! \begin{cases}
			\dfrac{1}{\max\{|\mathcal{N}_k|,|\mathcal{N}_i|\}},\ \text{if}\ i \in \mathcal{N}_k \slash\{k\}\\
			1-\sum_{i \in \mathcal{N}_k}c_{ik}, \ \text{if}\ i=k \\
			0, \  \text{otherwise}
		\end{cases}$\\
		\hline
	\end{tabularx}	
\end{table}

When~\eqref{eq:atc_dlms1} and~\eqref{eq:atc_dlms2} are performed in this order at each iteration, this corresponds to a configuration known in the literature as \emph{adapt-then-combine} (ATC)\mbox{\cite{Sayed_Networks2014,Sayed_capitulo2014,Sayed_Proc2014,lopes2008diffusion,cattivelli2009diffusion}}. The order of these steps could be reversed, resulting in the \textit{combine-then-adapt} (CTA) configuration. In this paper, we focus on the ATC protocol, but the results can extended to the CTA dLMS.

\section{Theoretical Analysis} \label{sec:analysis}

In our analysis, we are especially interested in the NMSD, a commonly adopted performance indicator, defined as~\cite{Sayed_Networks2014}
\begin{equation} \label{eq:nmsd}
	\text{NMSD}(n) \triangleq 	\frac{1}{V} \sum_{k=1}^V \text{MSD}_k(n),
\end{equation}
in which $\text{MSD}_k$ is the mean-square deviation at node $k$, given~by
\begin{equation} \label{eq:msd_k}
	\text{MSD}_k(n) \triangleq \E\{\|\wtil_k(n)\|^2\},
\end{equation}
where we have introduced the weight-error vector
\begin{equation} \label{eq:w_til}
	\widetilde{\w}_k(n) \triangleq \wo - \w_k(n).
\end{equation}
For the clarity of the exposition, it is convenient to introduce the quantities
\begin{equation} \label{eq:def_beta}
	\beta_{ij}(n) \triangleq \E\{\wtil^{\rm T}_i(n)\wtil_j(n)\}
\end{equation}
for $i=1,\cdots,V$ and $j = 1,\cdots, V$. Otherwise, the notation could become overloaded. It is worth noting that
\begin{equation} \label{eq:beta_kk}
	\beta_{kk}(n)  =\E\{\|\wtil_k(n)\|^2\} = \text{MSD}_k(n).
\end{equation}

We then introduce the matrix $\BM(n)$ such that $[\BM(n)]_{ij} \!=\! \beta_{ij}(n)$, i.e.,
\begin{equation} \label{eq:def_B}
	\BM(n) = \begin{bmatrix}
		\beta_{11}(n) & \beta_{12}(n) & \cdots & \beta_{1V}(n)\\
		\beta_{21}(n) &  \beta_{22}(n) & \cdots  & \beta_{2V}(n)\\
		\vdots & \vdots & \ddots & \vdots \\
		\beta_{V1}(n) & \beta_{V2}(n) & \cdots & \beta_{VV}(n)
	\end{bmatrix},
\end{equation}
which allows us to recast the NMSD as
\begin{equation} \label{eq:NMSD_Trace}
	\text{NMSD}(n) = \dfrac{1}{V} \Tr\{\BM(n)\}.
\end{equation}
Furthermore, defining the $V^2 \times 1$ vector
\begin{equation} \label{eq:def_beta_vetor}
	\boldsymbol{\beta}(n) \triangleq \vect\{\BM(n)\} = [	\beta_{11}(n)\ \beta_{21}(n)\ \cdots\ \beta_{VV}(n)]^{\rm T},
\end{equation}
and recalling that
\begin{equation} \label{eq:trace_property}
	\Tr\{\mathbf{M}_1\mathbf{M}_2\} = \vect\{\mathbf{M}_1\}^{\rm T}\vect\{\mathbf{M}_2\}
\end{equation}
for any arbitrary matrices $\mathbf{M}_1$ and $\mathbf{M}_2$ of compatible dimensions,~\eqref{eq:NMSD_Trace} can be written as
\begin{equation} \label{eq:NMSD_b}
	\text{NMSD}(n) = \dfrac{1}{V} \bM^{\rm T} \boldsymbol{\beta}(n),
\end{equation}
where we have defined $\bM \triangleq \vect\{\IM_V\}$.

Resuming our analysis, by subtracting both sides of~\eqref{eq:atc_dlms1} from $\wo$, and replacing~\eqref{eq:dk} and~\eqref{eq:ek} in the resulting equation, after some algebraic manipulations, we can write
\begin{equation} \label{eq:atc_dlms1_til}
\begin{aligned}
\widetilde{\boldsymbol{\psi}}_k(n) &= [\IM_M - \mu_k\zeta_k(n) \uM_k(n) \uM^{\rm T}_k(n)]\wtil_k(n-1)\\
& - \mu\zeta_k(n) \uM_k(n) v_k(n),
\end{aligned}
\end{equation}
where we have introduced
\begin{equation} \label{eq:psi_til}
	\widetilde{\boldsymbol{\psi}}_k(n) \triangleq \wo - \boldsymbol{\psi}_k(n).
\end{equation}

On the other hand, from~\eqref{eq:atc_dlms2}, we observe that
\begin{equation} \label{eq:atc_dlms2_til}
\wtil_k(n) = \sum_{i \in \mathcal{N}_k} c_{ik}\widetilde{\boldsymbol{\psi}}_i(n).
\end{equation}

If we multiply both sides of~\eqref{eq:atc_dlms2_til} by $\wtil_k^{\rm T}(n)$ from the left, and use~\eqref{eq:atc_dlms2} again, we obtain after some algebra

\begin{equation} \label{eq:atc_dlms2_til2}
\|\wtil_k(n)\|^2 = \sum_{i \in \mathcal{N}_k}\sum_{j \in \mathcal{N}_k}c_{ik}c_{jk}\widetilde{\boldsymbol{\psi}}_j^{\rm T}(n)\widetilde{\boldsymbol{\psi}}_i(n).
\end{equation}
Replacing~\eqref{eq:atc_dlms1_til} in~\eqref{eq:atc_dlms2_til2}, we obtain
\begin{equation} \label{eq:norma_wtil}
	\begin{aligned}
		&\|\wtil_k(n)\|^2 = \sum_{i \in \mathcal{N}_k}\sum_{j \in \mathcal{N}_k}c_{ik}c_{jk}\\
		& \!\!\cdot\! \Big\{\![\IM_M \!-\! \mu_j\zeta_j(n) \uM_j(n) \uM^{\rm T}_j(n)]\wtil_j(n\!-\!1) \\
		& - \mu_j\zeta_j(n) \uM_j(n) v_j(n)\!\Big\}^{\! \rm T}\\
		& \cdot \Big\{\![\IM_M \!-\! \mu_i \zeta_i(n) \uM_i(n) \uM^{\rm T}_i(n)]\wtil_i(n\!-\!1) \\
		& \!-\! \mu_i \zeta_i(n) \uM_i(n) v_i(n)\!\Big\}.
	\end{aligned}
\end{equation}

To examine the MSD of node $k$, we need to take the expectations from both sides of~\eqref{eq:norma_wtil}. At this point, we make a few assumptions to make the analysis more tractable, {all of which are common in the related literature~\cite{Sayed_Networks2014,Sayed_capitulo2014,Sayed_Proc2014,SayedL2008,Haykin_AFT2014,CapituloVitor}:}
\begin{itemize}
\item[\textbf{A1}.] All the nodes in the network employ the same step size, i.e., $\mu_1 = \cdots = \mu_V = \mu > 0;$
\item[\textbf{A2}.] The weight error vectors $\wtil_i(n-1)$ are statistically independent of $\uM_j(n)$ for any pair $i$ and $j$. This is a multi-agent version of the independence theory, a common assumption in the adaptive filtering literature~\cite{SayedL2008,Haykin_AFT2014};
\item[\textbf{A3}.] The measurement noise $v_k(n)$ is zero-mean with variance $\sigma_{v_k}^2$, independent and identically distributed (iid), and independent from any other variable for $k=1,\cdots,V$;
\item[\textbf{A4}.] The input signals are zero-mean and white Gaussian with variance $\sigma_{u_1}^2 = \cdots = \sigma_{u_V}^2 = \sigma_u^2 > 0$. In other words, the autocorrelation matrices $\RM_{u_k} \triangleq \E\{\uM_k(n)\uM_k^{\rm T}(n)\}$, $k=1,\cdots,V$ are the same, and are proportional to the identity matrix, i.e., $\RM_{u_1} = \cdots = \RM_{u_V} = \sigma_u^2 \IM_M$;
\item[\textbf{A5}.] For every node $k$, $\zeta_k(n)$ is independent from any other variable, and drawn from a Bernoulli distribution, such that $\zeta_k(n) \!=\! 1$ with probability $p_{\zeta}$ and  $\zeta_k(n) \!=\! 0$ with probability $1 \!-\! p_{\zeta}$. We remark that we are assuming that $p_{\zeta}$ is the same for every node in the network. Furthermore, for any pair of distinct nodes, $\zeta_i(n)$ and $\zeta_j(n)$, $i \neq j$, are statistically independent from each other;
\item[\textbf{A6}.] At any time instant $n$, $u_i(n)$ is statistically independent from $u_j(n)$ for any pair of nodes $i$ and~$j$, $i \neq j$.
\end{itemize}

With these assumptions at hand, we can continue with our analysis. For the sake of brevity, we shall omit here the intermediate steps and focus on the main results obtained from~\eqref{eq:norma_wtil}. These results are justified in detail in Appendix~\ref{sec:proof}. For the scenario described, using~\eqref{eq:ck}, we can obtain
\begin{multline}  \label{eq:recursion_beta_kk}
			\beta_{kk}(n) \! =\! \theta \sum_{i=1}^V c_{ik}^2\beta_{ii}(n-1)\\
	 + \tau \sum_{j=1}^V\sum_{\substack{\ell=1\\ \ell \neq j}}^V \!c_{jk}c_{\ell k}\beta_{j \ell}(n-1) + \mu^2  p_{\zeta} M \sigma_u^2 \sum_{q=1}^V c_{qk}^2 \sigma_{v_q}^2,	 
\end{multline}
where for the sake of compactness we have introduced
\begin{equation} \label{eq:theta}
\theta \triangleq 1 - 2 \mu  p_{\zeta} \sigma_u^2 + \mu^2  p_{\zeta} \sigma_u^4(M+2)
\end{equation}
and
\begin{equation} \label{eq:tau}
	\tau \triangleq 1 - 2 \mu  p_{\zeta} \sigma_u^2 + \mu^2  p_{\zeta}^2 \sigma_u^4.
\end{equation}

Hence, we can see that $\text{MSD}_k(n)$ depends on the MSD of its neighbors at the previous iteration, as well as on the trace of the cross correlation matrices between $\wtil_j(n-1)$ and \mbox{$\wtil_{\ell}(n-1)$}, i.e., $\beta_{j \ell}(n-1) \!=\! \E\{\wtil_j^{\rm T}(n-1)\wtil_{\ell}(n-1)\}$, for every pair of nodes $j$ and $\ell$ in the neighborhood of node $k$. The impact of each of these terms on the behavior of node $k$ depends on $\mu$, $\sigma_u^2$, $c_{ik}$ for $i \! \in \! \mathcal{N}_k$ (and, therefore, on the network topology), and, in the case of the MSD's of neighboring nodes, on the filter length $M$. Finally, the noise variance in the neighborhood also influences directly the MSD of node $k$.

From~\eqref{eq:recursion_beta_kk}, it becomes evident that we also need to study how the trace of the cross-correlation matrix of $\wtil_j(n)$ and $\wtil_{\ell}(n)$, with $j \neq \ell$, evolves over time. Again, we focus on the main result and leave the details for the Appendix~\ref{sec:proof}. Using assumptions \textbf{A1} to \textbf{A6}, we can obtain the following recursion:
\begin{multline} \label{eq:recursion_beta_ij}
		\beta_{j\ell}(n)  = \theta \sum_{t=1}^V c_{tj}c_{t\ell}  \beta_{tt}(n-1)\\
+\! \tau \!  \sum^{V}_{r = 1}\!\sum^{V}_{\substack{s = 1\\ s \neq r}}\! c_{rj} c_{s \ell} \beta_{rs}(n-1)
\!+\! \mu^2  p_{\zeta} M \sigma_u^2 \! \sum_{z = 1}^V\! c_{zj}c_{z\ell} \sigma_{v_z}^2\!.
\end{multline}

{Next, we analyze in Sec.~\ref{sec:non} the special case of the non-cooperative approach, as it is more straightforward and enables us to draw some qualitative conclusions about this type of scenario. Then, in Sec.~\ref{sec:general}, we resume our analysis of Eqs.~\eqref{eq:recursion_beta_kk} and~\eqref{eq:recursion_beta_ij} for the general case. Later on, in Sec.~\ref{sec:approx}, we derive an approximate model for the cooperative strategies that will provide us with valuable insights.}

\subsection{The Non-Cooperative Case} \label{sec:non}

For the non-cooperative case, the analysis is straightforward. Since in this case we have $c_{kk} \!=\! 1$ for any $k\!=\!1,\cdots,V$, and $c_{ik} \!=\! 0$ if $i \!\neq\! k$, we can recast~\eqref{eq:recursion_beta_kk} as
\begin{equation} \label{eq:beta_kk_nc}
	\beta_{kk}(n) = \theta \beta_{kk}(n-1) + \mu^2 p_\zeta M \sigma_u^2 \sigma_{v_k}^2.
\end{equation}
Assuming $\beta_{kk}(0) \!=\! \|\wo\|^2$, by recursively applying~\eqref{eq:beta_kk_nc} we get
\begin{equation} \label{eq:beta_kk_nc2}
	\beta_{kk}(n) = \theta^n \|\wo\|^2 + \mu^2 p_\zeta M \sigma_u^2 \sigma_{v_k}^2 \sum_{n_i=0}^{n-1} \theta^{n_i}.
\end{equation}
{Assuming that $|\theta| < 1$, we have that $\theta^n$ fades to zero as $n \rightarrow \infty$. At this point, it is worth noting that}
	\begin{equation} \label{eq:comparacao1}
{\tau = (1 - \mu  p_{\zeta} \sigma_u^2)^2 \geq 0}
	\end{equation}
	and that
	\begin{equation} \label{eq:comparacao2}
\theta = \tau + \mu^2 p_{\zeta}\sigma_u^4(M+2 - p_{\zeta}) \geq \tau,
	\end{equation}
where the equality only occurs if $p_\zeta = 0$, in which case $\theta = \tau = 1$. Hence, assuming $p_\zeta > 0$, we notice that $\theta < 1$ if, and only if
	\begin{equation} \label{eq:condicao_mu}
		0 < \mu  < \dfrac{2}{(M+2)\sigma_u^2},
	\end{equation}
	 where we have incorporated the fact that $\mu > 0$. {In this case, we can write $\sum_{n_i = 0}^{n-1} \theta^{n_i} = \frac{1- \theta^n}{1- \theta}$. Thus,} considering~\eqref{eq:nmsd} and~\eqref{eq:beta_kk}, by applying some algebraic manipulations to~\eqref{eq:beta_kk_nc2}, we can write the NMSD as
\boxedeqn{ \label{eq:NMSD_teorico_nc}
	\text{NMSD}_{\text{nc}}(n) = (\|\wo\|^2-\chi_{\text{nc}}) \theta^n + \chi_{\text{nc}},
}
with
\begin{equation} \label{eq:chi_nc}
	\chi_{\text{nc}} \triangleq \dfrac{\mu M}{2 - \mu \sigma_u^2(M+2)}\cdot \dfrac{\sum_{k=1}^V \sigma_{v_k}^2}{V}.
\end{equation} 
{Taking the limit $\lim_{n \rightarrow \infty} \text{NMSD}_{\text{nc}}(\infty)$ in~\eqref{eq:NMSD_teorico_nc} yields}
\boxedeqn{\label{eq:NMSD_teorico_ss_theta}
	\text{NMSD}_{\text{nc}}(\infty) = \chi_{\text{nc}}.
}
{Therefore, we can clearly see that $\chi_{\text{nc}}$ given by~\eqref{eq:chi_nc} represents the steady-state value of the NMSD for the non-cooperative strategy. It is interesting to notice that $p_\zeta$ does not appear in~\eqref{eq:chi_nc}. Thus, the sampling probability does not affect the steady-state NMSD of the algorithm in the non-cooperative approach whatsoever, so long as $p_\zeta > 0$. If $p_\zeta = 0$ were chosen, we would obtain $\theta = 1$, and, from~\eqref{eq:beta_kk_nc}, we would get $\beta_kk(n) = \|\wo\|^2$ for every iteration $n$. This is reasonable, since in this case the nodes would never acquire any information on the optimal system.}

{There are a couple more things to notice from the previous analysis. Firstly, we remark that~\eqref{eq:chi_nc} agrees with existing results in the literature for the steady-state NMSD of non-cooperative networks when all the nodes are sampled and employ sufficiently small step sizes~\cite{Sayed_Networks2014}. Moreover, taking into account that, for a single LMS filter, it is a well-known result that its MSD can be approximated by~\cite{SayedL2008,Haykin_AFT2014}}
\begin{equation} \label{eq:chi_lms}
{\chi_{\text{LMS}} \triangleq \dfrac{\mu M\sigma_v^2}{2 - \mu \sigma_u^2(M+2)}}
\end{equation} 	
{for sufficiently small step sizes, we see from~\eqref{eq:chi_nc} that the steady-state NMSD of the non-cooperative approach is simply the average of the MSD's of $V$ individual LMS filters working isolated from one another, with each filter $k$ subjected to a certain noise power $\sigma_{v_k}^2$. This conclusion makes sense, since there is no cooperation between the nodes. Lastly, we remark that~\eqref{eq:condicao_mu} is simply the condition for the stability of an LMS filter in the mean~\cite{SayedL2008,Haykin_AFT2014}. Therefore, we can interpret~\eqref{eq:condicao_mu} as follows: so long as $p_\zeta > 0$, if we pick a step size $\mu$ that leads to the individual stability of each filter in the network, the network as a whole will be stable, regardless of the value of $p_\zeta$.}

{Finally, we notice that in~\eqref{eq:NMSD_teorico_nc} the term $(\|\wo\|^2 - \chi_{\text{nc}})\theta^n$ decays exponentially along the iterations. Assuming that this term is dominant during the transient phase in comparison with $\chi_{\text{nc}}$, we conclude that the closer $\theta$ is to unity, the slower the convergence rate. From~\eqref{eq:theta}, we can clearly see that}
\begin{equation} \label{eq:lim_theta}
{\lim_{p_{\zeta} \rightarrow 0^+} \theta = 1.}
\end{equation}
{This indicates that the lower the sampling probability, the slower the convergence.}

{From the previous discussion, we can summarize the effects of sampling on the behavior of non-cooperative networks as in the following result.}

\begin{theorem}[Non-cooperative networks]
{In the case of the non-cooperative networks, the stability of dLMS is ensured if~\eqref{eq:condicao_mu} holds and $0 < p_\zeta \leq 1$. Under these conditions, the lower the sampling probability $p_\zeta$, the slower the convergence rate. Moreover, the steady-state NMSD is completely unaffected by $p_\zeta$. This result follows as a direct consequence of Eqs.~\eqref{eq:NMSD_teorico_nc}--\eqref{eq:NMSD_teorico_ss_theta} and~\eqref{eq:lim_theta}.}
\end{theorem}

\subsection{The General Case} \label{sec:general}

Let us now resume our analysis for a more general case, encompassing the cooperative strategies as well. From~\eqref{eq:recursion_beta_kk} and~\eqref{eq:recursion_beta_ij}, we observe that $\beta_{kk}(n)$ can be seen as a linear combination of the $\beta_{ii}(n-1)$ and $\beta_{j \ell}(n-1)$, plus the constant term $\mu^2 M \sigma_u^2 \sum_{q \in \mathcal{N}_k} c_{qk}^2 \sigma_{v_q}^2$. This becomes clear if we expand the first two summations in~\eqref{eq:recursion_beta_kk}, i.e.,
\begin{equation} \label{eq:recursion_beta_kk2}
	\begin{aligned}
		&\beta_{kk}(n) \! =\! \theta c_{1k}^2\beta_{11}(n\!-\!1) +  \cdots + \theta c_{Vk}^2\beta_{VV}(n\!-\!1) \\
		& \!+ \! \tau c_{1k}c_{2k}\beta_{1 2}(n\!-\!1) \!+\! \cdots \!+\! \tau c_{Vk}c_{(V\!-\!1)k}\beta_{V(V\!-\!1)}(n\!-\!1)\\
		& \!+\! \mu^2  p_{\zeta} M \sigma_u^2 \sum_{q=1}^V c_{qk}^2 \sigma_{v_q}^2,
	\end{aligned}
\end{equation}
Analogously, the same can be said about $\beta_{j \ell}(n)$ based on~\eqref{eq:recursion_beta_ij}. Hence, we should be able to write
\begin{equation} \label{eq:beta}
	\boldsymbol{\beta}(n) = \boldsymbol{\Phi}\boldsymbol{\beta}(n-1) + \mu^2  p_{\zeta} M \sigma_u^2 \boldsymbol{\sigma},
\end{equation}
where $\boldsymbol{\Phi}$ is a matrix whose $k$-th row determines how exactly each $\beta_{ij}(n-1)$ influences the corresponding term in the current iteration, and $\boldsymbol{\sigma}$ is a vector that aggregates the information from the network topology and noise variance from the constant terms that appear in~\eqref{eq:recursion_beta_kk} and~\eqref{eq:recursion_beta_ij}.

Let us now aggregate the combination weights into a $V \times V$ matrix $\CM$, such that $[\CM]_{ij} \!=\! c_{ij}$. Similarly, let us collect the noise variances in a $V \times V$ diagonal matrix $\RM_v$, such that its $k$-th element is equal to $\sigma_{v_k}^2$, i.e.,
\begin{equation}
	\RM_v = \begin{bmatrix}
		\sigma_{v_1}^2 & 0 & \cdots & 0\\
		0 & \sigma_{v_2}^2 & \cdots & 0\\
		\vdots & \vdots & \ddots & 0\\
		0 & 0 & \cdots & \sigma_{v_V}^2
	\end{bmatrix}.
\end{equation}

In this case, we observe that

\begin{equation} \label{eq:sigma}
\!\!\!\!\CM \RM_v\! \CM^{\rm T} \!\!=\!\!\! \arraycolsep0pt\begin{bmatrix} 
	\displaystyle\sum_{k=1}^V \! \! c_{k1}^2 \sigma_{v_k}^2 & \displaystyle\sum_{k=1}^V \! \! c_{k 1} c_{k 2} \sigma_{v_k}^2 & \cdots & \displaystyle\sum_{k=1}^V \! \! c_{k 1} c_{k V} \sigma_{v_k}^2\\
	\displaystyle\sum_{k=1}^V \! \! c_{k 2} c_{k 1} \sigma_{v_k}^2 & \displaystyle\sum_{k=1}^V \! \! c_{k 2}^2 \sigma_{v_k}^2 & \cdots & \displaystyle\sum_{k=1}^V \! \! c_{k 2} c_{k V} \sigma_{v_k}^2\\
	\vdots & \vdots & \ddots & \vdots\\
	\displaystyle\sum_{k=1}^V \! \! c_{k V} c_{k 1} \sigma_{v_k}^2 & \displaystyle\sum_{k=1}^V \! \! c_{k V}c_{k 2} \sigma_{v_k}^2 & \cdots  & \displaystyle\sum_{k=1}^V \! \! c_{k V}^2 \sigma_{v_k}^2\\
\end{bmatrix}\!\!\!.\!\!\!
\end{equation}
Then, we may write the $V^2 \times 1$ vector $\boldsymbol{\sigma}$ in~\eqref{eq:beta} as
\begin{equation}
	\boldsymbol{\sigma} = \vect\{\CM \RM_v\! \CM^{\rm T}\} = \begin{bmatrix}
		\sum_{i=1}^V c_{i1}^2 \sigma_{v_i}^2\\
		\sum_{i=1}^V c_{i2}c_{i1} \sigma_{v_i}^2\\
		\vdots\\
		\sum_{i=1}^V c_{iV}^2 \sigma_{v_i}^2
	\end{bmatrix}.
\end{equation}

As for the matrix $\boldsymbol{\Phi}$,  from~\eqref{eq:recursion_beta_kk} and~\eqref{eq:recursion_beta_ij} we obtain that
\begin{equation} \label{eq:phi}
\boldsymbol{\Phi} = \boldsymbol{\Omega} \odot \boldsymbol{\Gamma},
\end{equation}
where we have introduced
\begin{equation} \label{eq:gamma}
	\boldsymbol{\Gamma} \triangleq (\CM \otimes \CM)^{\rm T}
\end{equation}
and
\begin{equation} \label{eq:omega0}
	\boldsymbol{\Omega} = [\boldsymbol{\Omega}_1\ \boldsymbol{\Omega}_2 \ \cdots \ \boldsymbol{\Omega}_V],
\end{equation}
in which $\boldsymbol{\Omega}_i$ is a $V^2 \times V$ matrix whose elements in the $i$-th column are all equal to $\theta$, and whose other elements are all equal to $\tau$, i.e.
\begin{equation} \label{eq:omega_i}
\boldsymbol{\Omega}_i = \begin{blockarray}{ccccccc}
	& & & \makebox[0pt]{$i$\text{-th column}} & & & \\
	& & & \makebox[0pt]{$\downarrow$} & & & \\
	\begin{block}{[ccccccc]}
		\tau    & \cdots &   \tau   &   \theta  &   \tau    & \cdots &   \tau    \\
		\tau    & \cdots &   \tau   &   \theta  &   \tau    & \cdots &   \tau    \\
		\vdots & \ddots & \vdots & \vdots & \vdots & \ddots & \vdots\\
		\tau    & \cdots &   \tau   &   \theta  &   \tau    & \cdots &   \tau    \\				
	\end{block}
	& & & & \hspace*{-1.2cm} \makebox[0pt]{$\underbrace{\rule{14em}{0pt}}_{\normalsize V \ \text{columns}}$} & & \\
\end{blockarray}.
\end{equation}
More details on the matrices  $\boldsymbol{\Phi}$ and $\boldsymbol{\Omega}$ are discussed in Appendix~\ref{sec:phi}, but are skipped here for the sake of brevity. Interestingly, the matrix $\boldsymbol{\Phi}$ results from the element-wise multiplication of the matrices $\boldsymbol{\Gamma}$ and $\boldsymbol{\Omega}$. The matrix $\boldsymbol{\Gamma}$ is related to the combination weights and, consequently, to the combination step. The matrix $\boldsymbol{\Omega}$, on its turn, is related to the adaptation step through $\theta$ and $\tau$.

With Eqs.~\eqref{eq:sigma} to~\eqref{eq:omega_i} at hand, we can continue with the analysis of Eq.~\eqref{eq:beta}. Considering an initial condition $\boldsymbol{\beta}_0 = \boldsymbol{\beta}(0)$, the recursive application of~\eqref{eq:beta} leads to
\begin{equation} \label{eq:beta2}
	\boldsymbol{\beta}(n) = \boldsymbol{\Phi}^n\boldsymbol{\beta}_0 + \mu^2  p_{\zeta} M \sigma_u^2 \sum_{n_i=0}^{n-1}\boldsymbol{\Phi}^{n_i} \boldsymbol{\sigma}.
\end{equation}
If the algorithm is initialized with $\w_k(0) = \mathbf{0}_{M}$ for every node $k$, we have that $\wtil_k(0) = \wo$. Thus, for any $i$ and $j$, we have that $\beta_{ij}(0) = \E\{\wtil^{\rm T}_i(0)\wtil_j(0)\} = \E\{\wo^{\rm T}\wo\} = \|\wo\|^2$, and, consequently,
\begin{equation} \label{eq:beta0}
\boldsymbol{\beta}_0 = \|\wo\|^2 \mathbf{1}_{V^2}.
\end{equation}

Hence, replacing~\eqref{eq:beta0} in~\eqref{eq:beta2} and observing that $\sum_{n_i=0}^{n-1}\Phi^{n_i}=[\IM_{V^2}-\Phi]^{-1}[\IM_{V^2}-\Phi^n]$, we obtain
\begin{equation} \label{eq:beta3}
	\begin{aligned}
	\!\!\boldsymbol{\beta}(n) &\!=\! \|\wo\|^2 \boldsymbol{\Phi}^n\mathbf{1}_{V^2} \\
	&\!+\! \mu^2  p_{\zeta} M \sigma_u^2 [\IM_{V^2} \!-\! \boldsymbol{\Phi}]^{-1}[\IM_{V^2} \!-\! \boldsymbol{\Phi}^n]\boldsymbol{\sigma}.
	\end{aligned}
\end{equation}
Thus, considering~\eqref{eq:beta3} and~\eqref{eq:NMSD_b}, we can write
\boxedeqn{\label{eq:NMSD_teorico}
	\begin{aligned}
		&\text{NMSD}(n) \!=\! \dfrac{1}{V}\bigg\{\|\wo\|^2 \bM^{\rm T}\boldsymbol{\Phi}^n\mathbf{1}_{V^2} \\
		&\!+\! \mu^2  p_{\zeta}M \sigma_u^2 \bM^{\rm T}[\IM_{V^2} \!-\! \boldsymbol{\Phi}]^{-1}[\IM_{V^2} \!-\! \boldsymbol{\Phi}^n]\boldsymbol{\sigma}\!\bigg\}.
\end{aligned}}

It is worth noting that although $\wo$ appears in~\eqref{eq:beta3} and~\eqref{eq:NMSD_teorico}, we do not need to know it beforehand. Instead, we only need its norm. This is usually not a problem, since the norm of the optimal system can be adjusted by using adaptive gain control. Furthermore, we remark that $\beta_{ij}(n) = \beta_{ji}(n)$, which means that the matrix $\BM$ defined in~\eqref{eq:def_B} is symmetric. Thus, the vector $\boldsymbol{\beta}(n)$ given by~\eqref{eq:def_beta} has $V(V-1)/2$ duplicated entries. Although we could remove these elements from our model to make it more efficient from a computational perspective, and make the appropriate modifications where needed, we opted not to make this change for the clarity of the exposition. This is due to the fact that we are not primarily concerned with the computational complexity of the proposed model, and we believe that the formulation adopted is more convenient for the calculations. However, we would like to reinforce that this change is possible, and can reduce the amount of computations significantly, especially if $V$ is large. Lastly, it is interesting to notice that we can obtain the theoretical MSD of the LMS algorithm~\cite{SayedL2008,Haykin_AFT2014,CapituloVitor} as a special case of~\eqref{eq:NMSD_teorico}. In this situation, we have that $V = 1$, and $\mathbf{b}$, $\boldsymbol{\sigma}$, $\GAM$, $\boldsymbol{\Omega}$ and $\boldsymbol{\Phi}$ degenerate into $\mathbf{b} = 1$, $\boldsymbol{\sigma} = \sigma_v^2$, $\GAM = 1$, $\boldsymbol{\Omega} = \theta$ and $\boldsymbol{\Phi} = \theta$, respectively. Replacing these results in~\eqref{eq:NMSD_teorico}, we obtain
\begin{equation*}
\text{MSD}(n) = (\|\wo\|^2-\chi_{\text{LMS}}) \theta^n + \chi_{\text{LMS}},
\end{equation*}
with $\chi_{\text{LMS}}$ given by~\eqref{eq:chi_lms}.

From~\eqref{eq:beta3} and~\eqref{eq:NMSD_teorico}, we can see that the stability of the network in the mean-squared sense is ensured if $\lim_{n \rightarrow \infty} \boldsymbol{\Phi}^n = \mathbf{0}_{V^2 \times V^2}$, which occurs if, and only if,~\cite{Meyer2000}
\begin{equation} \label{eq:rho}
\rho(\boldsymbol{\Phi}) < 1,
\end{equation}
where $\rho(\cdot)$ denotes the spectral radius of a matrix, i.e., the maximum absolute value of its eigenvalues.

If~\eqref{eq:rho} holds, by taking the limit of~\eqref{eq:NMSD_teorico} as $n \rightarrow \infty$, we conclude that the steady-state NMSD is given by
\boxedeqn{
\label{eq:NMSD_teorico_ss}
\text{NMSD}(\infty) = \dfrac{\mu^2 p_\zeta M \sigma_u^2}{V} \bM^{\rm T}[\IM_{V^2} \!-\! \boldsymbol{\Phi}]^{-1}\boldsymbol{\sigma}.	
}

As will become clear in Sec.~\ref{sec:simulation}, the model described by~\eqref{eq:NMSD_teorico} can be very accurate. However, it still does not allow us to draw many qualitative conclusions about the behavior of the diffusion algorithm. Thus, further approximations may come in handy to aid us in this task, as shown next.


%

\subsection{An Approximate Model for the Cooperative Strategies} \label{sec:approx}

Since the columns of the matrix $\boldsymbol{\Omega}$ are filled by either $\tau$ or $\theta$, we could adopt $\boldsymbol{\Omega} \approx \tau \mathbf{1}_{V^2 \times V^2}$
or $\boldsymbol{\Omega} \approx \theta \mathbf{1}_{V^2 \times V^2}$ as an approximation. Making these replacements in~\eqref{eq:phi} leads to $\boldsymbol{\Phi} \approx \tau \GAM$ or $\boldsymbol{\Phi} \approx \theta \GAM$, respectively. {Due to~\eqref{eq:comparacao2}},
the second approximation tends to be more conservative. Replacing it in~\eqref{eq:rho}, and using the fact that for any real scalar $\alpha$ and matrix $\mathbf{M}$, $\rho(\alpha \mathbf{M}) = |\alpha|\rho(\mathbf{M})$, we conclude that
\begin{equation} \label{eq:rho2}
	\theta \cdot  \rho(\boldsymbol{\Gamma}) < 1,
\end{equation}
where we used the fact that $\theta > 0$. 

Due to~\eqref{eq:ck}, $\CM$ is a left-stochastic matrix, i.e., a matrix whose entries are all non-negative, and whose columns add up to one. Consequently, $\CM \otimes \CM$ is also a left-stochastic matrix. By transposing it, we conclude that $\boldsymbol{\Gamma}$ is a right-stochastic matrix, i.e., all of its entries are non-negative, and its rows add up to one. One interesting property of such matrices is that their spectral radius is always equal to one~\cite{Meyer2000,blondel2005convergence}. Thus, the condition established by~\eqref{eq:rho2} can be recast as simply $\theta< 1$. Replacing~\eqref{eq:theta} in~\eqref{eq:rho2} and assuming that $p_\zeta > 0$, after some algebra we {get~\eqref{eq:condicao_mu}}. Hence, we {can observe} that {our previous conclusion that} if $\mu$ lies within a certain range, the sampling probability $p_\zeta$ does not affect the stability of the algorithm, so long as $p_\zeta > 0$, {holds for the general case, and not just for the non-cooperative approach}. We remark that~\eqref{eq:condicao_mu} corresponds to ensuring that each individual filter in the non-cooperative scheme is stable. It is a well-known fact in the adaptive diffusion networks literature that, if every individual node is stable, the stability of the network as a whole in a cooperative scenario is also ensured~\cite{Sayed_Networks2014,Sayed_capitulo2014,Sayed_Proc2014}. However, {we remark that}~\eqref{eq:condicao_mu} was obtained considering a worst-case scenario, in which $\boldsymbol{\Phi} \!=\! \theta \GAM$. In practice,~\eqref{eq:condicao_mu} is not strictly necessary to ensure the stability of the algorithm {if a cooperative strategy is adopted}. In {these cases}, greater step sizes may be employed without making the algorithm unstable. In this case, it will be shown in Sec.~\ref{sec:simulation} that the sampling of the nodes may actually be beneficial to the stability. In other words, in the worst-case scenario, sampling does not hinder the stability of the algorithm, and, in general, it may improve~it.

Furthermore, simulation results show that, when the nodes do cooperate, the approximation $\boldsymbol{\Phi} \approx \tau \GAM$ leads to more accurate predictions than $\boldsymbol{\Phi} \approx \theta \GAM$. Thus, adopting the first approximation for the cooperative strategies, we obtain from~\eqref{eq:beta3} 
\begin{equation} \label{eq:NMSD_teorico_tau0}
	\begin{aligned}
		&\text{NMSD}(n) \!=\! \dfrac{1}{V}\bigg\{\|\wo\|^2 \bM^{\rm T}\tau^n\GAM^n\mathbf{1}_{V^2} \\
		&\!+\! \mu^2  \!p_{\zeta}M \sigma_u^2 \bM^{\rm T}[\IM_{V^2} \!-\! \boldsymbol{\Phi}]^{-1}[\IM_{V^2} \!-\! \tau^n\GAM^n]\boldsymbol{\sigma}\!\bigg\}\!.
	\end{aligned}
\end{equation}
We remark that the result of the multiplication $\boldsymbol{\Gamma}^n\mathbf{1}_{V^2}$ is a column vector whose $i$-th element is the sum of the elements of the $i$-th row of the matrix $\boldsymbol{\Gamma}^n$. Since the product of right-stochastic matrices is also right-stochastic~\cite{blondel2005convergence}, $\GAM^n$ is right-stochastic, from which we conclude that $\boldsymbol{\Gamma}^n\mathbf{1}_{V^2} = \mathbf{1}_{V^2}$.
Thus,~\eqref{eq:NMSD_teorico_tau0} can be recast as

\begin{equation} \label{eq:NMSD_teorico_tau1}
	\begin{aligned}
		&\text{NMSD}(n) \!=\! \dfrac{1}{V}\bigg\{\|\wo\|^2 \tau^n \bM^{\rm T}\mathbf{1}_{V^2} \\
		&\!+\! \mu^2  \!p_{\zeta}M \sigma_u^2 \bM^{\rm T}[\IM_{V^2} \!-\! \boldsymbol{\Phi}]^{-1}[\IM_{V^2} \!-\! \tau^n\GAM^n]\boldsymbol{\sigma}\!\bigg\}\!.
	\end{aligned}
\end{equation}
Using the fact that $\bM^{\rm T}\mathbf{1}_{V^2} = V$, we thus conclude that for the cooperative strategies, the NMSD is well approximated by
\boxedeqn{\label{eq:NMSD_teorico_tau}
	\begin{aligned}
	\text{NMSD}_{\tau}(n) &= \|\wo\|^2 \tau^n + \dfrac{\mu^2 p_{\zeta} M \sigma_u^2}{V}  \\
	&\!\!\!\!\!\!\!\! \cdot \bM^{\rm T} [\IM_{V^2} \!-\! \tau\boldsymbol{\Gamma}]^{-1}[\IM_{V^2} \!-\! \tau^n\boldsymbol{\Gamma}^n]\boldsymbol{\sigma}.
\end{aligned}
}


{Analogously to what we observed in Sec.~\ref{sec:non} about~\eqref{eq:NMSD_teorico_nc}, the first term in~\eqref{eq:NMSD_teorico_tau} decays exponentially along the iterations with $\tau^n$}. Assuming {once again} that {this} term {is} dominant during the transient phase, we conclude that the closer that {$\tau$ is} to unity, the slower the convergence rate. From~\eqref{eq:theta} and~\eqref{eq:tau}, we {observe} that
\begin{equation} \label{eq:lim_tau}
	\lim_{p_{\zeta} \rightarrow 0^+} \tau = 1.
\end{equation}
{Hence, much like in the non-cooperative case of Sec.~\ref{sec:non}}, the lower the sampling probability, the slower the convergence {for the cooperative strategies. We remark that this result} is in accordance with Fig.~\ref{fig:intro}.

Finally, for the steady state, {assuming that $\tau < 1$ and taking the limit when $n \rightarrow \infty$ in~\eqref{eq:NMSD_teorico_tau} yields}
\boxedeqn{\label{eq:NMSD_teorico_ss_tau}
	\text{NMSD}_{\tau}(\infty) \!=\!  \dfrac{\mu^2 p_{\zeta} M \sigma_u^2}{V} \bM^{\rm T}[\IM_{V^2} \!-\! \tau \boldsymbol{\Gamma}]^{-1}\boldsymbol{\sigma}.}

Eq.~\eqref{eq:NMSD_teorico_ss_tau} clearly depends on the matrix $\boldsymbol{\Gamma}$, and, therefore, on the network topology and combination rule adopted. It is not straightforward to extract conclusions from~\eqref{eq:NMSD_teorico_ss_tau} and~\eqref{eq:NMSD_teorico_ss_theta} for any arbitrary topology without calculating them explicitly. However, it may be interesting to analyze a particular case. Let us consider a network topology represented by a complete graph, i.e., one in which every pair of nodes is directly connected by an edge. A graph with this topology and $V$ nodes is usually denoted by $K_V$ in the literature. An example with $V=8$ nodes is depicted in Fig.~\ref{fig:K8}. 
\begin{figure}[htb!]
	\centering
	\includegraphics[width = 0.6\columnwidth]{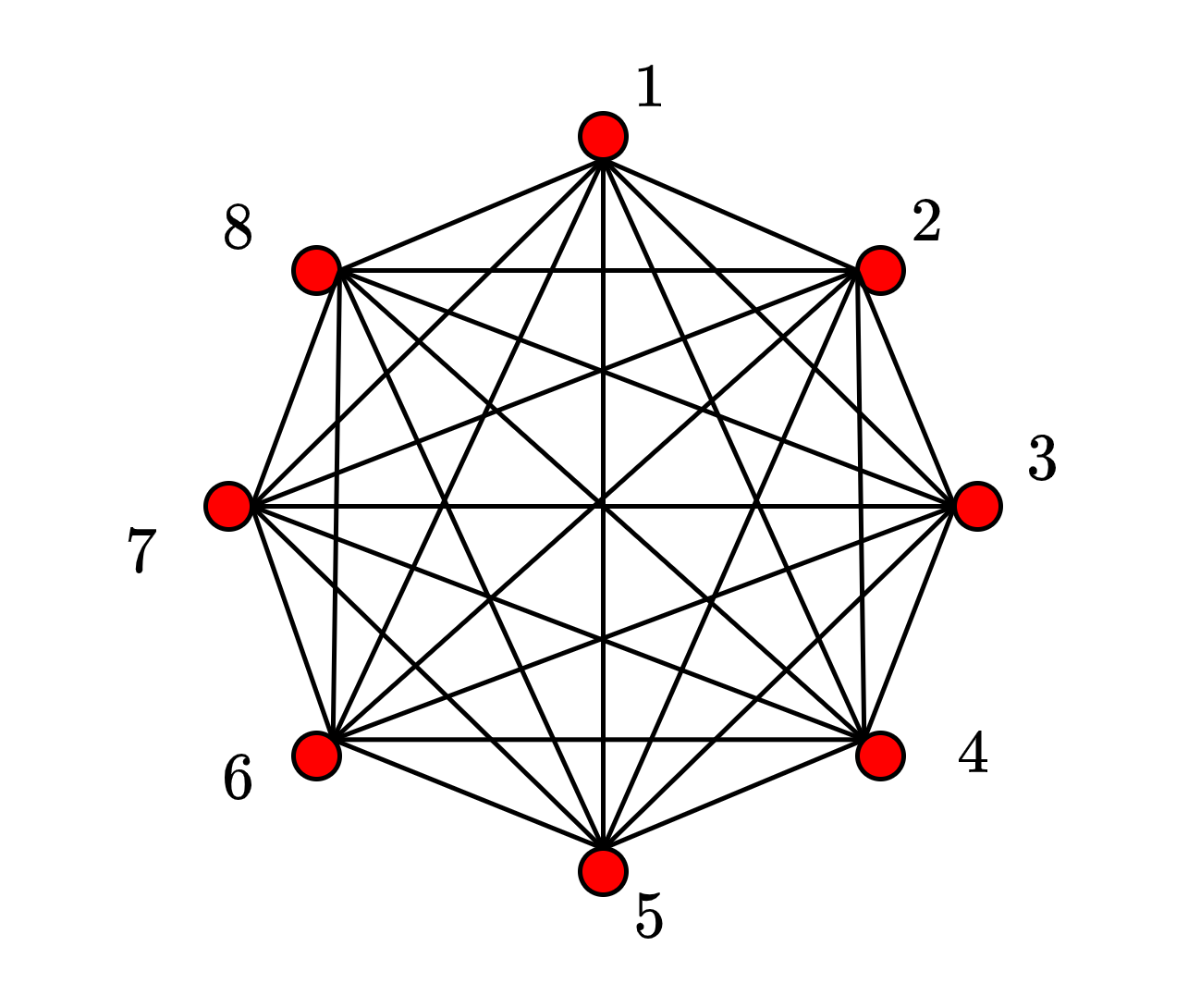}
	\caption{\protect\label{fig:K8} A network arranged according to the $K_8$ topology.}
\end{figure}

For the $K_V$ topology, the Uniform and Metropolis weights coincide, and lead to $\CM_{K_V} = \frac{1}{V} \mathbf{1}_{V \times V}$ and $\boldsymbol{\Gamma}_{K_V} = \frac{1}{V^2}\mathbf{1}_{V^2 \times V^2}$, where the index $K_V$ is adopted to refer to this type of topology with Uniform or Metropolis weights. For the aforementioned matrix $\boldsymbol{\Gamma}_{K_V}$, from~\eqref{eq:NMSD_teorico_ss_tau}, we get
\begin{equation} \label{eq:NMSD_teorico_ss_tau_kv}
\text{NMSD}_{\tau_{K_V}}(\infty) = \chi_{K_V} \triangleq  \dfrac{\mu M}{2 - \mu p_{\zeta} \sigma_u^2}\dfrac{\sum_{k=1}^V \sigma_{v_k}^2}{V^2}.
\end{equation}
This result is derived in Appendix~\ref{sec:kv_ss}. Comparing~\eqref{eq:chi_nc},~\eqref{eq:NMSD_teorico_ss_theta}, and~\eqref{eq:NMSD_teorico_ss_tau_kv}, we observe that the sampling probability does not affect the steady-state performance of the non-cooperative strategy, but does influence the steady-state NMSD of the cooperative schemes. From~\eqref{eq:NMSD_teorico_ss_tau_kv}, we get
\begin{equation} \label{eq:limit}
\dfrac{\mu M}{2}\dfrac{\sum_{k=1}^V \sigma_{v_k}^2}{V^2} < \chi_{K_V} \leq \dfrac{\mu M}{2 - \mu \sigma_u^2}\dfrac{\sum_{k=1}^V \sigma_{v_k}^2}{V^2},
\end{equation}
i.e., as we reduce the value of $p_{\zeta}$, $\chi_{K_V}$ decreases as well. The first inequality is obtained by taking the limit of $\chi_{K_V}$ as $p_\zeta \rightarrow 0^+$. This observation is in accordance with the results from Fig.~\ref{fig:intro}. This reduction in the steady-state NMSD should be more significant for relatively large values of $\mu$ and $\sigma_u^2$. If $\mu \sigma_u^2 \ll 2$, however, the impact of sampling becomes negligible. We remark that~\eqref{eq:chi_nc} and~\eqref{eq:NMSD_teorico_ss_tau_kv} agree with the analysis of~\cite{Sayed_Networks2014,Sayed_capitulo2014,Sayed_Proc2014} for the ATC dLMS algorithm with every node sampled and small step sizes, if we consider $p_\zeta \!=\! 1$ in the latter. {Lastly, it is worth noting that~\eqref{eq:NMSD_teorico_ss_tau_kv} only holds for $p_\zeta > 0$. This is because, in order to obtain this result, we assumed in our calculations that $\tau < 1$, which can only be true if $p_\zeta > 0$, as can be seen from~\eqref{eq:tau}. If we consider $p_\zeta = 0$, we would get $\tau = 1$ and therefore obtain from~\eqref{eq:NMSD_teorico_ss_tau} that $\text{NMSD}_\tau(n) = \|\wo\|^2$ for every iteration $n$, which is in accordance with our expectations.}

{We can summarize the results of the analysis for the cooperative networks as in the following result.}

\begin{theorem}[Cooperative networks]
{In the case of the cooperative networks, the stability of dLMS is ensured in the mean-squared sense if~\eqref{eq:rho} holds, which can only occur if $p_\zeta > 0$. If the sampling probability is different from zero,~\eqref{eq:condicao_mu} is a sufficient but not strictly necessary condition for stability of cooperative networks. Furthermore, if the algorithm is stable, then the lower the sampling probability $p_\zeta$, the slower the convergence rate, much like in the non-cooperative case. However, in contrast with the non-cooperative approach, in cooperative schemes the steady-state NMSD decreases as we reduce $p_\zeta$. This follows as a consequence of Eqs.~\eqref{eq:rho} and~\eqref{eq:NMSD_teorico_tau}--\eqref{eq:NMSD_teorico_ss_tau}, and is better visualized from the approximate model given by~\eqref{eq:NMSD_teorico_ss_tau_kv}.}
\end{theorem}

{Comparing Results 1 and~2, therefore, we can see that in both the non-cooperative and cooperative schemes, the convergence rate is deteriorated as we reduce $p_\zeta$. However, it is only in the latter case that the steady-state NMSD decreases.} One possible interpretation is as follows. The adaptation step is the process through which the algorithm acquires knowledge about its environment. For this reason, it is particularly relevant when there is little knowledge about the optimal system -- e.g., during the transient phase. Thus, it makes sense that by not sampling the nodes in the transient phase, the convergence rate should deteriorate. However, the adaptation step also introduces noise into the algorithm, since it involves the acquisition of the desired signal, which is corrupted by it. In steady state, the algorithm does not gain enough information from the adaptation step to continue to improve its estimate of the optimal system, but is affected by the noise that it injects. The step size directly influences the impact of the measurement noise on the algorithms, since it multiplies $d_k(n)$, and, therefore, $v_k(n)$ in~\eqref{eq:atc_dlms1}. Thus, the greater the $\mu$, the more the noise will affect the behavior of the algorithm. Conversely, if $\mu$ is small, this effect is restricted. Following this line of reasoning, if we cease to sample some of the nodes, there is less noise entering the algorithm. In a non-cooperative scheme, if a node is not sampled, its estimate remains fixed until its sampling is resumed. Hence, there is no reason why the {steady-state performance should be affected} by sampling less nodes, which is predicted by our theoretical model. However, in the cooperative schemes, this is changed by the existence of the combination step. Indeed,~\eqref{eq:NMSD_teorico_ss_tau_kv} shows that we should expect some {decrease} in the {steady-state NMSD} in this scenario, even if slightly. The theoretical model attributes this to the parameter $\tau$, which is related to the cooperation between nodes in~\eqref{eq:recursion_beta_kk}, since it determines how $\E\{\wtil_{\ell}^{\rm T}(n-1)\wtil_j(n-1)\}$, for $\ell$ and $j$ in the neighborhood  of node $k$, will affect $\E\{\|\wtil_k(n)\|^2\}$. The model also shows that when $\mu$ is large, the {effects of sampling less nodes on the steady-state NMSD should be more noticeable }, which supports the idea of the step size as a factor that determines the impact of the measurement noise in the algorithm. Interestingly, the diagnosis that the adaptation step injects noise in the algorithm, and that the combination step tends to remove it, has been raised in, e.g.,~\cite{tiglea2020low,tiglea2022adaptive,lee2018data},  but lacked formal theoretical support, until now.

Lastly, it may be interesting to compare the steady-state NMSD achieved by the algorithm with a certain step size $\mu$ and sampling probability $p_\zeta$, and the one obtained by utilizing $\mu^\prime = \mu p_\zeta$ and maintaining all nodes sampled. Denoting these quantities by $\text{NMSD}_{\text{async.}}(\infty)$ and $\text{NMSD}_{\text{sync.}}(\infty)$, respectively, we notice from~\eqref{eq:NMSD_teorico_ss_tau_kv} that, for the topology of Fig.~\ref{fig:K8}, we get
\begin{equation}
	\text{NMSD}_{\text{sync.}}(\infty) = p_{\zeta }\text{NMSD}_{\text{async.}}(\infty).
\end{equation}
This is in accordance with~\cite{zhao2014asynchronous3}, in which it was noticed that the steady-state NMSD of the synchronous networks should be less than or equal to that of the asynchronous one, if an adjusted step size is adopted taking $p_\zeta$ into consideration. In other words, if we adjust the step size, the network with all nodes sampled should outperform the one with random sampling, as it will present approximately the same convergence rate, with a lower steady-state NMSD. However, for a fixed step size, our conclusion that sampling less nodes improves the steady-state performance remains valid. This is relevant because we could devise algorithms that keep the sampling of the nodes during the transient phase, so as to maintain a fast convergence rate, and that sample less nodes in steady state. By doing so, we simultaneously reduce the computational cost and the NMSD in steady state while keeping $\mu$ fixed. That is the goal of, e.g., the algorithms of~\cite{tiglea2020low,tiglea2022adaptive}.

\section{Computational Cost Reduction} \label{sec:computational}

In this section, we seek to estimate the effects of sampling on the expected computational cost of the dLMS algorithm. For brevity, we focus on the number of multiplications per iteration, but a similar analysis could be done for the additions.

We begin by noticing that each sampled node $k$ needs to perform $M(2 + |\mathcal{N}_k|) + 1$ multiplications per iteration, assuming that static combination weights are employed. This is summarized in Table~\ref{tab:sampled_nodes}, in which the number of multiplications required at each node $k$ per iteration  is denoted by $\otimes_k$ and detailed for each calculation required. We can see that $2M + 1$ multiplications are related to the adaptation step. Assuming that none of the operations associated with this step have to be performed at node $k$ if it is not sampled, we conclude that the total number of multiplications required at the iteration $n$ and node $k$ with random sampling can be estimated as
\begin{equation} \label{eq:mult_k}
\otimes_k(n) = \zeta_k(n) \cdot (2 M + 1) + M |\mathcal{N}_k|.
\end{equation}
Taking the expectations from both sides in~\eqref{eq:mult_k}, we obtain
\begin{equation} \label{eq:mult_k_expected}
	\E\{\otimes_k\} = p_\zeta(2 M + 1) + M |\mathcal{N}_k|,
\end{equation}
where we dropped the indication of the time instant $n$ since the right-hand side of~\eqref{eq:mult_k_expected} remains constant along the iterations.

\begin{table}[h!tb]
	\centering
	\caption{\protect\label{tab:sampled_nodes} Estimated number of multiplications per iteration at each sampled node $k$.}
	\begin{tabular}{|c||c|c|c|}
		\hline
		& Calculation  & Step & $\otimes_k$ \\
		\hline
		1& $y_k(n) = \mathbf{u}_k^{\rm T}\w_k(n-1)$ & Adapt. & $M$ \\
		\hline
		2& $e_k(n) = d_k(n) - y_k(n)$ & Adapt. & $0$ \\
		\hline
		3& $\mu \cdot e_k(n)$ & Adapt. & $1$ \\
		\hline
		4& $[\mu \cdot e_k(n)] \cdot \uM_k(n)$  & Adapt. & $M$ \\
		\hline							
		5& $\w_k(n\!-\!1)\!+\! \{[\mu \!\cdot\! e_k(n)] \!\cdot\! \uM_k(n)\}$  & Adapt. & $0$ \\
\hline
		6& $\sum_{i \in \mathcal{N}_k}c_{ik}\psiM_i(n)$  & Comb. & $M|\mathcal{N}_k|$ \\
\hline
\hline
\multicolumn{3}{|c|}{Total}& $M(2 + |\mathcal{N}_k|) + 1$\\
\hline
	\end{tabular}
\end{table}

Summing~\eqref{eq:mult_k_expected} for $k = 1, \cdots, V$, we can estimate the expected computational cost for the whole network as
\begin{equation} \label{eq:mult_expected}
	\E\{\otimes_{\text{total}}\} = Vp_\zeta(2 M + 1) + M \sum_{k=1}^V|\mathcal{N}_k|.
\end{equation}
By replacing $p_\zeta \!=\! 1$ in~\eqref{eq:mult_expected}, we obtain the number of multiplications required by the dLMS with every node sampled. Thus, denoting the number of multiplications saved per iteration due to the sampling by $\Delta \otimes_{\text{total}}$, we can thus estimate it as
\begin{equation} \label{eq:economy_mult}
	\E\{\Delta \otimes_{\text{total}}\} = V(2M+1)(1-p_\zeta)
\end{equation}
by subtracting~\eqref{eq:mult_expected} from the case with $p_\zeta = 1$. 

From~\eqref{eq:economy_mult}, we see that the smaller the $p_\zeta$, the greater the savings in computation, as expected. Moreover, for a given $p_\zeta$, the number of multiplications saved increases with $V$ and $M$.


\section{Simulation Results} \label{sec:simulation}

In this section, we present simulation results to validate the theoretical analysis. They were obtained over an average of 1000 independent realizations, considering the scenarios summarized in Table~\ref{tab:scenarios}. The Scenario 1 of this table corresponds to that of Fig.~\ref{fig:intro}. In every case, the coefficients of the optimal system $\wo$ are drawn from a uniform distribution in the range $[-1,1]$, and later normalized so that $\wo$ has unit norm. Moreover, the length of the adaptive filter is always equal to that of $\wo$. We consider the network topology presented in Fig.~\ref{fig:rede}, which was generated randomly. The input signal $u_k(n)$ and the measurement noise $v_k(n)$ follow Gaussian distributions with zero mean for each node $k$, with $\sigma_{u_k}^2 \!=\! \sigma_u^2 \!=\! 1$, whereas the noise variance $\sigma_{v_k}^2$ is drawn from a uniform distribution in the range $[0.001,0.01]$ for $k\!=\!1,\cdots,V$, as depicted in Fig.~\ref{fig:ruido}.

			\begin{figure}[htb!]
				\centering
				\begin{subfigure}[t]{0.49\columnwidth}
					\centering
						\includegraphics*[trim = 3.5cm 8cm 2.5cm 7.5cm, clip = true, width=1\columnwidth]{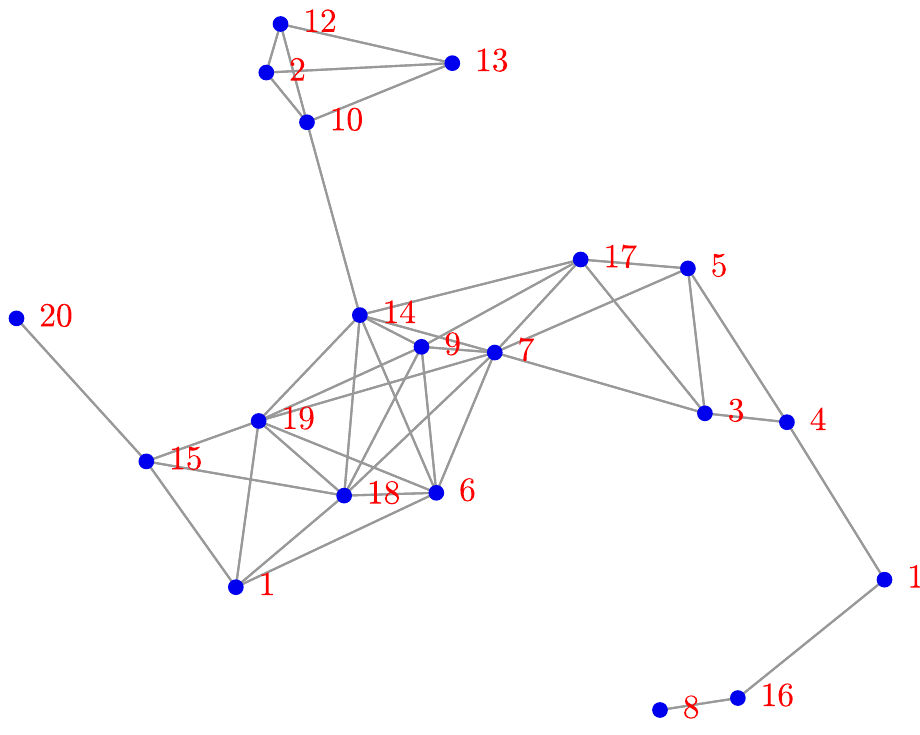}
						\caption{\protect\label{fig:rede}}
				\end{subfigure}
				\begin{subfigure}[t]{0.49\columnwidth}
					\centering
	\includegraphics*[trim = 0.1cm 0.2cm 6.7cm 0.1cm, clip = true, width=1\columnwidth]{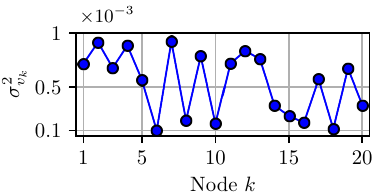}
	
	\caption{\protect\label{fig:ruido}}
				\end{subfigure}
	\caption{\protect\label{fig:network} (a) Network topology, and (b) noise variance profile considered in the simulations.}
\end{figure}


\begin{table}[h!tb]
	\centering
	\caption{\protect\label{tab:scenarios} List of scenarios considered in the simulations.}
	\begin{tabular}{|c|c|c|c|}
		\hline
		Scenario & $\mu$ & $M$  & Combination Rule\\
		\hline
		Scenario 1 & 0.1 & 10  & Metropolis\\
		\hline
		Scenario 2 & 0.01 & 10  & Metropolis\\
		\hline
				Scenario 3 & 0.02 & 100 & Uniform\\		
		\hline
		Scenario 4 & 0.01 & 10 & Non-Cooperative\\
\hline							
	\end{tabular}
\end{table}

This section is organized as follows. In Sec.~\ref{sec:transient}, we study the transient performance of the algorithm, in Sec.~\ref{sec:stability}, its stability, and in Sec.~\ref{sec:steady-state}, its steady-state NMSD.

\subsection{Transient Performance} \label{sec:transient}

In Fig.~\ref{fig:tres_cenarios}, we present the simulation results obtained in the Scenarios 1, 2, and 3 of Table~\ref{tab:scenarios}, considering $p_\zeta \in \{1, 0.5, 0.1\}$, and compare them to the theoretical models of Sec.~\ref{sec:analysis}. The sub-figures in the top row present a comparison with the more precise model of Eq.~\eqref{eq:NMSD_teorico}, whereas those in the bottom row show the comparison with the approximate model of Eq.~\eqref{eq:NMSD_teorico_tau}. Furthermore, each column of Fig.~\ref{fig:tres_cenarios} refers to a scenario, with Figs.~\ref{fig:tres_cenarios}  (a) and (b) presenting the results obtained in Scenario 1, Figs.~\ref{fig:tres_cenarios} (c) and (d) those from Scenario~2, and Figs.~\ref{fig:tres_cenarios} (e) and (f) those from Scenario 3.

From Figs.~\ref{fig:tres_cenarios} (a), (c), and (e), we can see that the simulation results match the theoretical curves very well for all three scenarios, and that the model accurately predicts the improvement in steady-state NMSD caused by the sampling of less nodes. Furthermore, comparing the results of Figs.~\ref{fig:tres_cenarios}~(c) and (e) with those of Fig.~\ref{fig:tres_cenarios} (a), we can observe that the difference in performance caused by the sampling is indeed more noticeable for larger step sizes, as expected. By comparing Figs.~\ref{fig:tres_cenarios}~(a) and~\ref{fig:tres_cenarios}~(b) can see that, in Scenario 1, the approximate model is less accurate than the one described by Eq.~\eqref{eq:NMSD_teorico}. This was expected, to a certain extent. The same can be said about Scenario 3, by comparing Figs.~\ref{fig:tres_cenarios}~(e) and (f). However, we observe from Figs.~\ref{fig:tres_cenarios}~(c) and (d) that, in Scenario 2, both models practically coincide. Hence, we can conclude that the approximate model of Eq.~\eqref{eq:NMSD_teorico_tau} tends to be more accurate for relatively low step sizes $\mu$ and filter lengths $M$, and is more affected by them than the model of Eq.~\eqref{eq:NMSD_teorico}.

\begin{figure*}[h!tb]
\centering
\includegraphics*[trim=0.15cm 0.2cm 0.15cm 0.1cm, clip=true,width=1\textwidth]{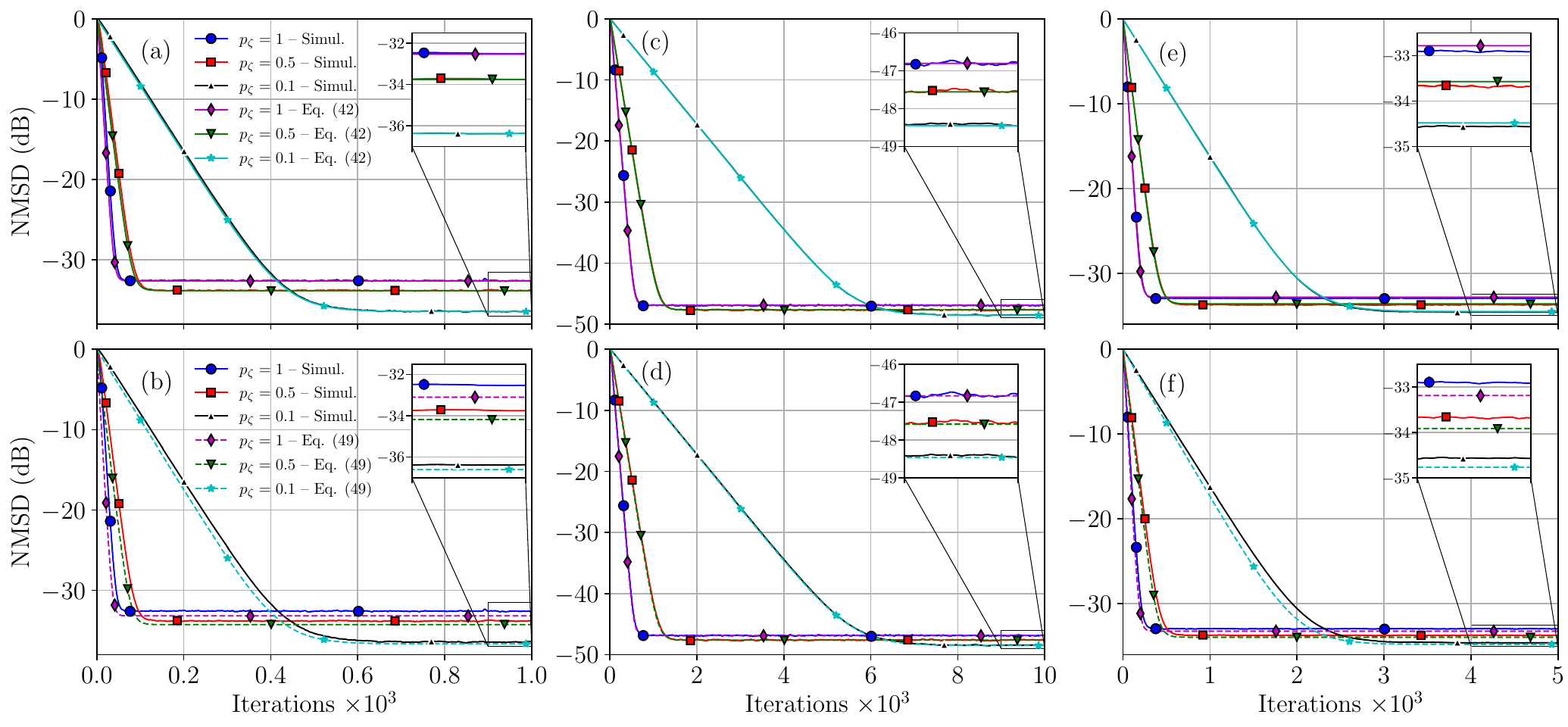}
\caption{\protect\label{fig:tres_cenarios} The sub-figures in the top row present a comparison between the simulation results and the model of Eq.~\eqref{eq:NMSD_teorico}, whereas the ones in the bottom row show a comparison with the model of Eq.~\eqref{eq:NMSD_teorico_tau}. The simulation results were obtained considering the Scenario 1 of Table~\ref{tab:scenarios} in sub-figures (a) and (b), Scenario 2 in (c) and (d), and Scenario~3 in (e) and (f).}
\end{figure*}



In order to examine the impact of sampling on the computational cost in this scenario, in Table~\ref{tab:cost_scenario1} we present the average number of multiplications required per iteration in the whole network for each $p_\zeta$ considered in the simulations for $M = 10$, as in Scenarios 1 and 2, and for $M = 100$, as in Scenario 3. We also present the number of multiplications saved per iteration in comparison with the case in which every node is sampled, and compare them to the results given by~\eqref{eq:economy_mult}. We can see that the average number of operations saved per iterations matches Eqs.~\eqref{eq:economy_mult}, which can be attributed to the high number of realizations and iterations considered in the computations. Furthermore, it is straightforward to see that the smaller the $p_\zeta$, the greater the savings in terms of computation, as expected. Moreover, for a fixed value of $p_\zeta$, the computational cost and the savings increase with $M$, as expected. From these experiments, we can summarize the effects of sampling as follows: smaller sampling probabilities $p_\zeta$ lead to lower steady-state NMSD and computational costs, at the expense of a deteriorated convergence rate. 

\begin{table}[h!tb]
	\centering
	\caption{\protect\label{tab:cost_scenario1} Average number of multiplications per iteration in the network for $p_\zeta \!\in\! \{0.1, 0.5, 1\}$ with $M \!=\! 10$ and $M \!=\! 100$.}
	\begin{tabular}{|c|cc|cc|cc|}
		\hline
		\multirow{2}{*}{$p_\zeta$} & \multicolumn{2}{c|}{$\otimes_{\text{total}}$} & \multicolumn{2}{c|}{$\Delta \otimes_{\text{total}}$} & \multicolumn{2}{c|}{Eq.~\eqref{eq:economy_mult}} \\ \cline{2-7} 
		& \multicolumn{1}{c|}{\!\!$M \!=\! 10$\!\!}   & \!\!$M \!=\! 100\!\!$   & \multicolumn{1}{c|}{\!\!$M \!= \!10$\!\!}       & \!\!$M = 100$\! \!     & \multicolumn{1}{c|}{\!\!$M \!=\! 10$\!\!}                & \!\!$M \!=\! 100$\!\!                \\ \hline
		1                          & \multicolumn{1}{c|}{$\approx\! 1460$}       & 14420       & \multicolumn{1}{c|}{0}              & 0              & \multicolumn{1}{c|}{0}                       & 0                        \\ \hline
		0.5                        & \multicolumn{1}{c|}{1250}       & 12410       & \multicolumn{1}{c|}{210}            & 2010           & \multicolumn{1}{c|}{210}                     & 2010                     \\ \hline
		0.1                        & \multicolumn{1}{c|}{1082}       & 10802       & \multicolumn{1}{c|}{378}            & 3618           & \multicolumn{1}{c|}{378}                     & 3618                     \\ \hline
	\end{tabular}
\end{table}

In the simulations of Fig.~\ref{fig:scenario3_exact}, we consider the Scenario~4 of Table~\ref{tab:scenarios}, and compare the simulation results to the theoretical model given by Eq.~\eqref{eq:NMSD_teorico_nc} for the non-cooperative scheme. Once again, the simulation results closely match the theoretical analysis. Furthermore, the simulations support the idea that the sampling probability does not affect the steady-state performance of the algorithm in the non-cooperative scheme, unlike what was observed in Fig.~\ref{fig:tres_cenarios} for the cooperative rules.

\begin{figure}[h!tb]
	\centering
	\includegraphics*[trim=0.15cm 0.2cm 0.15cm 0.1cm, clip=true,width=0.75\columnwidth]{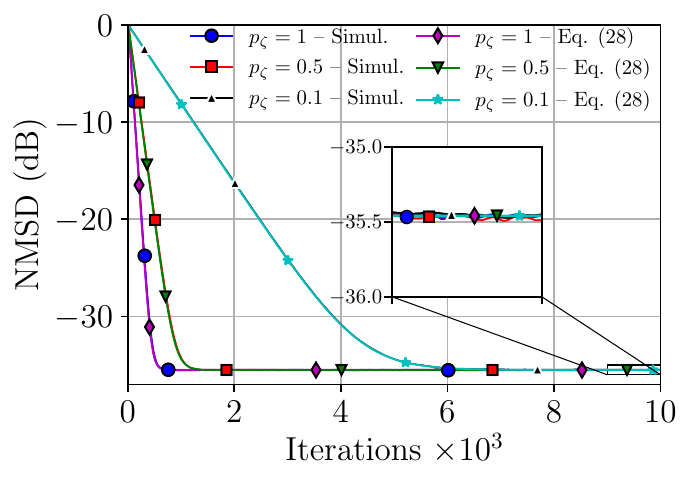}
	\caption{\protect\label{fig:scenario3_exact} Comparison between the simulation results and the model of Eq.~\eqref{eq:NMSD_teorico_nc} for $p_\zeta \!\in\! \{1, 0.5, 0.1\}$ in Scenario~4.}
\end{figure}

\subsection{Effects of Sampling on the Stability} \label{sec:stability}

From~\eqref{eq:condicao_mu}, we concluded that the sampling probability does not affect the stability of the algorithm so long as $\mu$ is sufficiently small and $p_\zeta \!>\! 0$. However,~\eqref{eq:condicao_mu} is not strictly necessary to ensure the stability of the algorithm in the mean-squared sense. For instance, the value of $\mu$ considered in the Scenario 3 of Table~\ref{tab:scenarios} does not satisfy~\eqref{eq:condicao_mu}, but still leads to the stability of the dLMS algorithm for the network of Fig.~\ref{fig:rede} with Uniform weights and $M \!=\! 100$. Under these conditions, one obtains $\rho(\boldsymbol{\Phi}) \approx 0.9628$ for $p_\zeta \!=\! 1$, which satisfies~\eqref{eq:rho} and thus ensures the convergence in the mean-squared sense. For $p_\zeta \!=\! 0.5$ and $p_\zeta \!=\! 0.1$, we get $\rho(\boldsymbol{\Phi}) \!<\! 1$ as well.

To verify if the sampling of the nodes influences the stability of the algorithm in the general case, we calculated the spectral radius of the matrix $\boldsymbol{\Phi}$ considering $M \!=\! 100$ and the three combination policies for the network of Fig.~\ref{fig:rede} with $\mu \!=\! 0.1$ and several values of $p_\zeta$. The results are shown in Fig.~\ref{fig:estabilidade}~(a), where we have highlighted with a red horizontal line the threshold $\rho(\boldsymbol{\Phi}) \!=\! 1$. We can see that, for all combination policies, the adoption of $p_\zeta \!=\! 0$ leads to $\rho(\boldsymbol{\Phi}) \!=\! 1$. This is expected, since in this case we get $\theta = \tau = 1$, and consequently $\boldsymbol{\Phi} \!=\! \GAM$, whose spectral radius is equal to one. Intuitively, this comes from the fact that the algorithm never acquires any knowledge on the optimal system if the nodes are never sampled. For the non-cooperative strategy, $\rho(\boldsymbol{\Phi})$ increases with $p_\zeta$, indicating that the algorithm is unstable for any sampling probability. For the Uniform and Metropolis combination policies, however, $\rho(\boldsymbol{\Phi})$ decreases up to a certain point with the increase of $p_\zeta$, and then begins to rise. Interestingly, for both policies, Fig.~\ref{fig:estabilidade}~(a) tells us that, under the conditions considered, the algorithm is unstable with all nodes sampled, but we can stabilize it by sampling less nodes. For the Uniform rule, we conclude from Fig.~\ref{fig:estabilidade}~(a) that the dLMS algorithm is stable for $p_\zeta \in ]0, 0.71]$ (approximately), whereas for the Metropolis rule the stability occurs for $p_\zeta \in ]0, 0.39]$. In order to verify these results, we ran the dLMS algorithm under the same circumstances considered in Fig.~\ref{fig:estabilidade}~(a) with different sampling probabilities in the range $[0.01, 1]$ for $200 \cdot 10^3$ iterations, which is more than necessary for the algorithm to achieve the steady state with $p_\zeta \!=\! 0.01$ and cooperative strategies. Then, utilizing the \code{isnan} and \code{isinf} functions of MATLAB\textregistered, we calculated the percentage of realizations in which the dLMS algorithm diverged at some iteration. The results are depicted in Fig.~\ref{fig:estabilidade}~(b). We can see that, for the non-cooperative strategy, the algorithm diverges in 100\% of the realizations for all values of $p_\zeta$ considered. For the Uniform and Metropolis rules, the percentage of realizations in which the algorithm diverges is initially zero, and increases steeply as $p_\zeta$ approaches the limit values of $p_\zeta \!=\! 0.71$ and $p_\zeta \!=\! 0.39$, respectively. For the former combination policy, the algorithm starts to diverge for $p_\zeta \!>\! 0.68$, whereas for the latter the first divergences occur for $p_\zeta \!>\! 0.41$. In both cases, by increasing $p_\zeta$ slightly further, the algorithm begins to diverge at some point in 100\% of the realizations. Therefore, the simulation results of Fig.~\ref{fig:estabilidade}~(b) support the theoretical findings of Fig.~\ref{fig:estabilidade}~(a). {It is worth noting that, although the Uniform rule leads to the stability of the algorithm for a wider range of $p_\zeta$ than the Metropolis rule in this case, this does not necessarily occur in all scenarios. For example, for the topology in Fig.~\ref{fig:K8}, the weights coincide for the two rules and therefore there is no difference between them in terms of the stability of the algorithm.}

\begin{figure}[h!tb]
	\centering
	\includegraphics*[trim=0.15cm 0.2cm 0.15cm 0.1cm, clip=true,width=0.8\columnwidth]{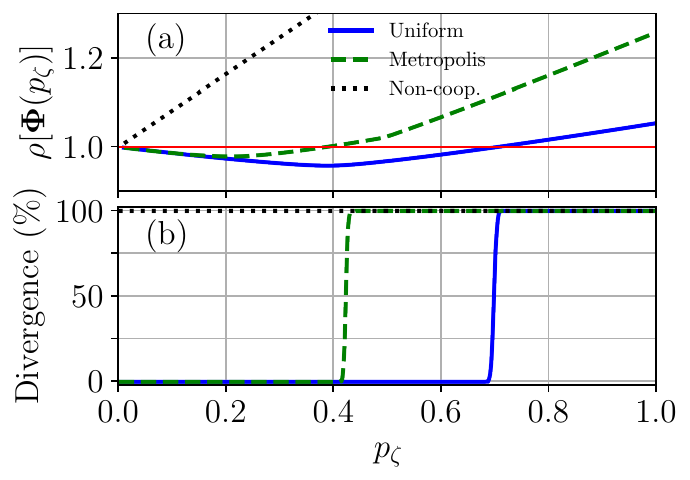}
	\caption{\protect\label{fig:estabilidade} (a) $\rho(\boldsymbol{\Phi})$ as a function of $p_\zeta$, and (b) percentage of realizations in which the dLMS diverged with $\mu \!=\! 0.1$ and $M\!=\!100$ for $p_\zeta \!\in\! [0.01, 1]$ with different combination policies.}
\end{figure}

\subsection{Steady-State Performance} \label{sec:steady-state}

Lastly, in order to verify Eqs.~\eqref{eq:NMSD_teorico_ss} and~\eqref{eq:NMSD_teorico_ss_tau} in detail, we ran the ATC dLMS for different values of $p_\zeta \in [0.1, 1]$, and calculated the average NMSD during the final 20\% iterations of each realization. The results are shown in Figs.~\ref{fig:regime} (a), (b), and (c) for Scenarios 1, 2, and 3, respectively. In each case, we set the total number of iterations $N$ so that the algorithms achieved the steady state before the end of each realization, resulting in $10^3 \leq N \leq 10^5$. In all scenarios, we observe that the steady-state NMSD drops continuously as we decrease $p_\zeta$. Moreover, the simulation results match~\eqref{eq:NMSD_teorico_ss} very closely in Scenarios 1 and 2. In Scenario 3, there is a discrepancy between the simulation results and the theoretical curve of 0.1 dB, on average. As for the model of Eq.~\eqref{eq:NMSD_teorico_ss_tau}, there is a difference of approximately 0.40 dB in comparison with the simulation results in Fig.~\ref{fig:regime} (a), on average. In Fig.~\ref{fig:regime}(c), this difference is of roughly 0.23 dB, whereas in Fig.~\ref{fig:regime}~(b) the curve practically overlaps with the simulation results.

\begin{figure}[h!tb]
	\centering
	\includegraphics*[trim=0.15cm 0.2cm 0.15cm 0.1cm, clip=true,width=0.77\columnwidth]{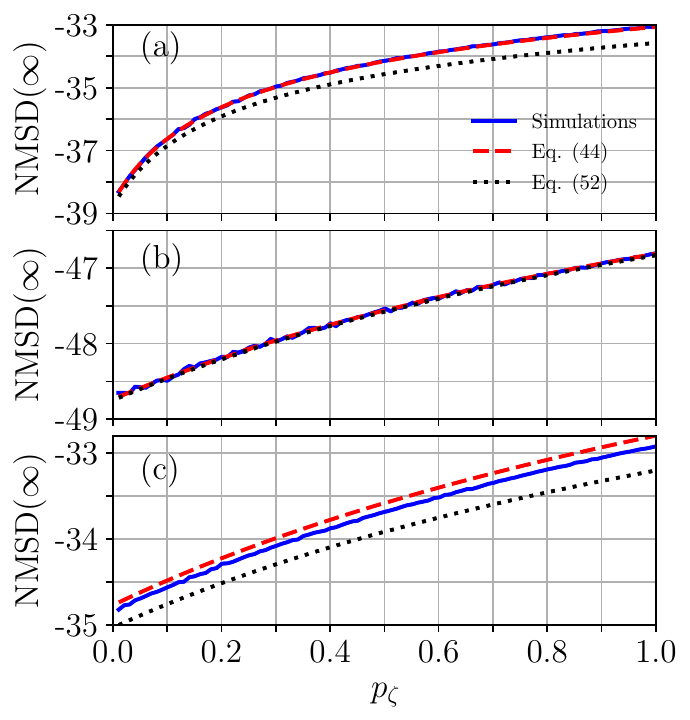}
	\caption{\protect\label{fig:regime} Steady-state NMSD for $p_\zeta \!\in\! [0.01, 1]$ in: (a) Scenario~1, (b) Scenario 2, and (c) Scenario 3.}
\end{figure}

\section{Conclusions} \label{sec:conclusions}

In this paper, we conducted a theoretical analysis on the ATC dLMS algorithm with random sampling in a stationary environment. It was shown that, so long as the sampling probability is greater than zero, the sampling does not hinder the stability of the algorithm in the worst-case scenario, and, in general, it may improve it. Furthermore, the theoretical model indicates that, for a fixed step size, the convergence rate of the algorithm deteriorates as we decrease the sampling probability, but the computational cost decreases and the steady-state {NMSD is slightly reduced}.  Finally, the analysis shows that this phenomenon is more noticeable when
the step size is fairly high. The simulations support the theoretical results obtained in the paper. In future works, we intend to extend the analysis to: i) non-Gaussian and/or colored input signals, {ii) to scenarios in which each node $k$ has its own distinct step size $\mu_k$, sampling probability $p_{\zeta_k}$, and autocorrelation matrix $\RM_{u_k}$ for the input vector $\uM_k(n)$, iii) to situations in which the optimal system varies over time, such as in random-walk scenarios, and iv) possibly to other solutions, such as the dNLMS and dRLS algorithms}. Finally, it may be interesting to investigate if this phenomenon also occurs in other types of algorithms, such as nonlinear solutions like the diffusion Kernel Least-Mean-Squares (dKLMS) algorithm~\cite{gao2015diffusion,chouvardas2016diffusion,shin2018distributed,bouboulis2018random}.

\appendices
\section{Deriving~\eqref{eq:recursion_beta_kk} and~\eqref{eq:recursion_beta_ij}} \label{sec:proof}

Taking the expectations from both sides of~\eqref{eq:norma_wtil} and using Assumption \textbf{A1}, we get
\begin{equation} \label{eq:norma_wtil_esperanca}
		\E\{\|\wtil_k(n)\|^2\} = \beta_{kk}(n) = \sum_{i \in \mathcal{N}_k}\sum_{j \in \mathcal{N}_k}c_{ik}c_{jk} x_{ji}(n),
\end{equation}
where we have defined
\begin{equation} \label{eq:norma_wtil_esperanca_x}
	\begin{aligned}
		&x_{ji}(n) \!\triangleq\!  \E\bigg\{\!\!\Big\{[\IM_M \!-\! \mu \zeta_j(n) \uM_j(n) \uM^{\rm T}_j(n)]\wtil_j(n\!-\!1)\\
		& - \mu \zeta_j(n) \uM_j(n) v_j(n)\!\Big\}^{\! \rm T}\\
		& \cdot\! \Big\{\![\IM_M \!-\! \mu  \zeta_i(n) \uM_i(n) \uM^{\rm T}_i(n)]\wtil_i(n\!-\!1) \\
		&- \mu  \zeta_i(n) \uM_i(n) v_i(n)\Big\}\!\!\bigg\}.
	\end{aligned}
\end{equation}

The analysis of~\eqref{eq:norma_wtil_esperanca_x} can be broken down into two cases: i) when $j = i$, and ii) when $j \neq i$. In the first situation, using \textbf{A3} and \textbf{A5}, and observing that $\E\{\zeta_i(n)\} = \E\{\zeta_i^2(n)\} = p_{\zeta}$,
we can write
\begin{equation} \label{eq:norma_wtil_esperanca_i_e_j_amostragem}
	\begin{aligned}
		&\!\!\!\! x_{ii}(n) \!=\! \E\{\wtil_i^{\rm T}(n\!-\!1)\wtil_i(n\!-\!1)\}\!\! \!\\
		& \!\!\!\!\!-\! 2\mu  p_{\zeta}\E\{\wtil_i^{\rm T}(n\!-\!1) \uM_i(n)\uM_i^{\rm T}(n)\wtil_i(n\!-\!1)\}\!\! \!\!\!\\
		&\!\!\!\!\!\!+\! \mu^2 p_{\zeta}\E\{\wtil_i^{\rm T}(n\!-\!1) \uM_i(n)\uM_i^{\rm T}(n)\uM_i(n)\uM_i^{\rm T}(n)\wtil_i(n\!-\!1)\}\!\! \!\!\\
		& \!\!\!\!\!+\! \mu^2 p_{\zeta}\sigma_{v_i}^2\E\{\uM_i^{\rm T}(n)\uM_i(n)\}.\!\! \!\!\!
	\end{aligned}
\end{equation}

Using Assumptions \textbf{A2} and \textbf{A4}, and following similar procedures to those used in the analysis of the MSD of the LMS algorithm, we may write (see pages 803--807 of~\cite{CapituloVitor})
\begin{equation}
\E\{\wtil_i^{\rm T}(n\!-\!1) \uM_i(n)\uM_i^{\rm T}(n)\wtil_i(n\!-\!1)\} =\sigma_u^2 \beta_{ii}(n-1)
\
\end{equation}
and
\begin{equation}
\begin{aligned}
&\E\{\wtil_i^{\rm T}(n\!-\!1) \uM_i(n)\uM_i^{\rm T}(n)\uM_i(n)\uM_i^{\rm T}(n)\wtil_i(n\!-\!1)\} =\\ &\sigma_u^{4}(M+2) \beta_{ii}(n-1).
\end{aligned}
\end{equation}
Thus,~\eqref{eq:norma_wtil_esperanca_i_e_j_amostragem} can be recast as
\begin{equation} \label{eq:norma_wtil_esperanca_i_e_j2}
 x_{ii}(n) \!=\! \theta \beta_{ii}(n-1) + \mu^2 p_{\zeta}M\sigma_u^2\sigma_{v_i}^2,
\end{equation}
with $\theta$ defined as in~\eqref{eq:theta}. Let us now analyze the case in which $j \neq i$. To make this distinction clearer, we shall replace the index $i$ by $\ell$ in this case. From \textbf{A5}, we can observe that
\begin{equation} \label{eq:esperanca_zeta}
	\E\{\zeta_{j}(n)\zeta_{\ell}(n)\} = \E\{\zeta_{j}(n)\}\E\{\zeta_{\ell}(n)\} = p_{\zeta}^2.
\end{equation}
Using~\eqref{eq:esperanca_zeta}, \textbf{A3} and \textbf{A5}, we can rewrite~\eqref{eq:norma_wtil_esperanca_x} for $\ell \neq j$ as
\begin{equation} \label{eq:norma_wtil_esperanca_i_neq_j_amostragem}
	\begin{aligned}
		&\! x_{j\ell}(n) \!=\! \E\{\wtil_j^{\rm T}(n\!-\!1)\wtil_{\ell}(n\!-\!1)\}\\
		& \!-\mu  p_{\zeta} \E\{\wtil_j^{\rm T}(n\!-\!1)\uM_j(n) \uM^{\rm T}_j(n)\wtil_{\ell}(n\!-\!1)\} \\
		&\! -\mu  p_{\zeta} \E\{\wtil_j^{\rm T}(n\!-\!1)\uM_{\ell}(n) \uM^{\rm T}_{\ell}(n)\wtil_{\ell}(n\!-\!1)\}\\
		&  \! + \! \mu^2  p_{\zeta}^2 \E\{\wtil_j^{\rm T}(n\!-\!1) \uM_j(n)\uM_j^{\rm T}(n)\uM_{\ell}(n)\uM_{\ell}^{\rm T}(n)\wtil_{\ell}(n\!-\!1)\}\!\!\!\!\!\!
	\end{aligned}
\end{equation}
Using \textbf{A2}, \textbf{A4}, and \textbf{A6}, from~\eqref{eq:norma_wtil_esperanca_i_neq_j_amostragem} we can write
\begin{equation}
\begin{aligned}
&\E\{\wtil_j^{\rm T}(n\!-\!1)\uM_j(n) \uM^{\rm T}_j(n)\wtil_{\ell}(n\!-\!1)\}\\
& = \E\{\wtil_j^{\rm T}(n\!-\!1)\uM_i(n) \uM^{\rm T}_i(n)\wtil_{\ell}(n\!-\!1)\}\\
& = \sigma_u^2 \beta_{j \ell}(n-1)
\end{aligned}
\end{equation}
for any pair of nodes $\ell$ and $j$, $\ell \neq j$. Furthermore, we notice that in the fourth-order moment that appears in~\eqref{eq:norma_wtil_esperanca_i_e_j_amostragem} is not present in~\eqref{eq:norma_wtil_esperanca_i_neq_j_amostragem}, and that we may write
\begin{equation}
\begin{aligned}
&\E\{\wtil_j^{\rm T}(n\!-\!1) \uM_j(n)\uM_j^{\rm T}(n)\uM_{\ell}(n)\uM_{\ell}^{\rm T}(n)\wtil_{\ell}(n\!-\!1)\}\\
& = \sigma_u^4 \beta_{j \ell}(n-1).
\end{aligned}
\end{equation}
Therefore, with $\tau$ defined as in~\eqref{eq:tau}, we can write
\begin{equation} \label{eq:norma_wtil_esperanca_i_neq_j2}
\!\!\!\! x_{j \ell}(n) \!=\! \tau \beta_{j \ell}(n-1),
\end{equation}
Thus, replacing~\eqref{eq:norma_wtil_esperanca_i_e_j2} and~\eqref{eq:norma_wtil_esperanca_i_neq_j2} in~\eqref{eq:norma_wtil_esperanca} leads to~\eqref{eq:recursion_beta_kk}.

As evidenced by~\eqref{eq:recursion_beta_kk} and~\eqref{eq:norma_wtil_esperanca_i_neq_j2}, we also need to obtain a recursion for $\E\{\wtil_j^{\rm T}(n-1)\wtil_{\ell}(n-1)\}$, for $j \neq \ell$, in order to analyze the evolution of the MSD of each node. Firstly, we should notice that we can rewrite $\wtil_j^{T}(n)\wtil_{\ell}(n)$ as a function of the local estimates, i.e.
\begin{equation} \label{eq:atc_dlms2_til_appendix}
\wtil_j^{\rm T}(n)\wtil_{\ell}(n) = \sum_{s \in \mathcal{N}_{\ell}}\sum_{r \in \mathcal{N}_j}c_{s \ell}c_{rj}\widetilde{\boldsymbol{\psi}}_r^{\rm T}(n)\widetilde{\boldsymbol{\psi}}_{s}(n).
\end{equation}

Replacing~\eqref{eq:atc_dlms1_til} in~\eqref{eq:atc_dlms2_til_appendix}, using Assumption \textbf{A1}, and taking the expectations from both sides, we arrive at
\begin{equation} \label{eq:norma_wjwi_esperanca}
\beta_{j \ell}(n) = \sum_{s \in \mathcal{N}_{\ell}}\sum_{r \in \mathcal{N}_j}c_{s \ell}c_{rj} x_{r s}(n).
\end{equation}
Similarly to what we did for~\eqref{eq:norma_wtil_esperanca}, the analysis of~\eqref{eq:norma_wjwi_esperanca} can be broken down into two cases: when $s = r = t$, and when $s \neq r$. In the first case, we have that $c_{tj} \neq 0$ and $c_{t \ell} \neq 0$ only if the node $t$ is in $\mathcal{N}_j\cap\mathcal{N}_{\ell}$. Thus, following an analogous procedure, we can write
\begin{equation} \label{eq:norma_wtil_esperanca_r_e_l2}
\beta_{tt}(n) \!=\! \mu^2 p_{\zeta}M\sigma_u^2\sigma_{v_t}^2 + \theta\beta_{tt}(n-1).
\end{equation}
For $s \neq r$, we can write
\begin{equation} \label{eq:norma_wtil_esperanca_r_neq_l}
\beta_{rs}(n) \!=\! \tau \beta_{rs}(n-1)
\end{equation}
Thus, replacing~\eqref{eq:norma_wtil_esperanca_r_e_l2} and~\eqref{eq:norma_wtil_esperanca_r_neq_l} in~\eqref{eq:norma_wjwi_esperanca}, we finally obtain~\eqref{eq:recursion_beta_ij}.

\section{On the matrix $\boldsymbol{\Phi}$} \label{sec:phi}

We begin by noting that~\eqref{eq:recursion_beta_ij} can be recast as
\begin{equation} 
	\begin{aligned}
		 \beta_{j\ell}(n)  &= \sum_{r=1}^V\sum_{s=1}^V c_{rj} c_{s \ell} [(\theta-\tau)\delta_{rs}+\tau]\beta_{rs}(n-1)\\
		&+ \mu^2 p_\zeta M \sigma_u^2\sum_{z = 1}^V c_{zj}c_{z\ell} \sigma_{v_z}^2,
	\end{aligned}
\end{equation}
for any arbitrary $j$ and $\ell$, by simply changing the order in which the elements are added. Thus, if $r = s$, $\beta_{rr}(n-1)$, which corresponds to the MSD of node $r$, is multiplied by $\theta$ and by $c_{rj}c_{r\ell}$. In contrast, if $r \neq s$, $\beta_{rs}(n-1)$ corresponds to the trace of the covariance matrix between $\wtil_r(n-1)$ and $\wtil_s(n-1)$, and is multiplied by $\tau$ and by $c_{rj}c_{s\ell}$. Thus, if we examine the vector $\boldsymbol{\beta}(n)$ in~\eqref{eq:def_beta_vetor}, we notice that it consists of $V$ elements between each pair of consecutive MSD's in the vector, and $V(V-1)$ elements related to cross-terms.

Thus, we conclude that, the matrix $\boldsymbol{\Phi}$ that appears in~\eqref{eq:beta} is a matrix that has $V$ columns filled with $\theta$, and between each pair of consecutive columns, there are $V$ columns filled with $\tau$. These columns are multiplied element-wise by the corresponding combination weights. As an example, let us consider a network formed by only two connected nodes. In this case, we have
\begin{equation}
\boldsymbol{\Phi} = \begin{bmatrix}
\theta c_{11}^2 & \tau c_{21}c_{11} & \tau c_{11}c_{21} & \theta c_{21}^2\\
\theta c_{12}c_{11} & \tau c_{22}c_{11} & \tau c_{12}c_{21} & \theta c_{22}c_{21}\\
\theta c_{11}c_{12} & \tau c_{21}c_{12} & \tau c_{11}c_{22} & \theta c_{21}c_{22}\\
\theta c_{12}^2 & \tau c_{22}c_{12} & \tau c_{12}c_{22} & \theta c_{22}^2
\end{bmatrix}.
\end{equation}

We should notice that there are $V=2$ columns that are related to the MSD's, and, in between them, we have also two columns, which are related to the covariances. If we focus on the combination weights, we can see that the matrix $\boldsymbol{\Phi}$ carries information from $(\CM^{\rm T}) \otimes (\CM^{\rm T}) \!=\! (\CM \otimes \CM)^{\rm T},$ where the equality follows from the properties of the Kronecker product. It also carries information from $\tau$ and $\theta$. Hence, we can see $\boldsymbol{\Phi}$ as the element-wise multiplication of two matrices, as in~\eqref{eq:phi}: $\boldsymbol{\Gamma}$, which is related to the combination weights as in~\eqref{eq:gamma}, and $\boldsymbol{\Omega}$, which is related to $\theta$ and $\tau$ as in~\eqref{eq:omega0}.

\section{Obtaining~\eqref{eq:NMSD_teorico_ss_tau_kv}} \label{sec:kv_ss}
Firstly, it is useful to note that
\begin{equation}
[\IM_{V^2} - \tau \boldsymbol{\Gamma}]^{-1} = \sum_{n_i=0}^{\infty} (\tau \boldsymbol{\Gamma})^{n_i} = \IM_{V^2} + \sum_{n_i=1}^{\infty} (\tau \boldsymbol{\Gamma})^{n_i}.
\end{equation} 
At this point, one useful property of the matrix $\boldsymbol{\Gamma}_{K_V} = \dfrac{1}{V^2}\mathbf{1}_{V^2 \times V^2}$ is that $\boldsymbol{\Gamma}_{K_V}^2 = \boldsymbol{\Gamma}_{K_V}$. Thus, we have that
\begin{equation} \label{eq:78}
[\IM_{V^2} - \tau \boldsymbol{\Gamma}_{K_V}]^{-1} = \IM_{V^2} + \dfrac{\tau}{(1- \tau)V^2}\mathbf{1}_{V^2 \times V^2}.
\end{equation}
Multiplying~\eqref{eq:78} from the left by $\bM^{\rm T}$ leads to
\begin{equation} \label{eq:79}
	 \bM^{\rm T}[\IM_{V^2} - \tau \boldsymbol{\Gamma}_{K_V}]^{-1} = \bM^{\rm T} + \dfrac{\tau}{(1- \tau)V} \mathbf{1}_{V^2}^{\rm T},
\end{equation}
By applying the inverse $\vect$ operator to the right-hand side of~\eqref{eq:79}, we obtain
\begin{equation} \label{eq:vec-1}
\vect^{-1}\left\{\bM^{\rm T} \!+\! \dfrac{\tau}{(1- \tau)V} \mathbf{1}_{V^2}\right\} \!=\! \IM_{V} \!+\! \dfrac{\tau}{(1- \tau)V} \mathbf{1}_{V \times V}.
\end{equation}
Thus, using~\eqref{eq:trace_property},~\eqref{eq:79}, and~\eqref{eq:vec-1}, and introducing $\xi \!\triangleq\! \bM^{\rm T}[\IM_{V^2} \!-\! \tau \boldsymbol{\Gamma}_{K_V}]^{-1}\vect\{\CM \RM_v \CM^{\rm T}\}$ for compactness, we can write
\begin{equation}
\begin{aligned}
& \xi \!=\! \Tr\left\{\left[\IM_{V} \!+\! \dfrac{\tau}{(1- \tau)V} \mathbf{1}_{V \times V}\right]\dfrac{1}{V^2}\mathbf{1}_{V \times V}\RM_v\mathbf{1}_{V \times V}\right\}\\
& = \dfrac{1}{V^2}\bigg(1+\dfrac{\tau}{1-\tau}\bigg)\Tr\{\mathbf{1}_{V \times V}\RM_v\mathbf{1}_{V \times V}\},
\end{aligned}
\end{equation}
where we used the fact that $\CM_{K_V} \!=\! \CM^{\rm T}_{K_V} \!=\! \frac{1}{V}\mathbf{1}_{V \times V}$ and that $\mathbf{1}_{V \times V}^2 \!=\! V \mathbf{1}_{V \times V}$. Furthermore, since $\Tr\{\mathbf{M}_1\mathbf{M}_2\mathbf{M}_3\} \!=\! \Tr\{\mathbf{M}_2\mathbf{M}_3\mathbf{M}_1\}$ for any arbitrary matrices $\mathbf{M}_1$, $\mathbf{M}_2$ and $\mathbf{M}_3$, we get
\begin{equation}
\Tr\{\mathbf{1}_{V \times V}\RM_v\mathbf{1}_{V \times V}\} = V \Tr\{\RM_v\mathbf{1}_{V \times V}\} = V \sum_{k=1}^V \sigma_{v_k}^2,
\end{equation}
where we took advantage from the fact that $\RM_v$ is a diagonal matrix. Thus, we obtain
\begin{equation} \label{eq:last}
\xi =  \dfrac{1}{V^2} \bigg(1+\dfrac{\tau}{1-\tau}\bigg) \cdot V \cdot \sum_{k=1}^V\sigma_{v_k}^2 = \dfrac{1}{1-\tau} \dfrac{\sum_{k=1}^V \sigma_{v_k}^2}{V}.
\end{equation}
Finally, replacing~\eqref{eq:last} in~\eqref{eq:NMSD_teorico_ss_tau} leads to~\eqref{eq:NMSD_teorico_ss_tau_kv}.


\end{document}